\documentclass{aa}  

\usepackage{graphicx}
\usepackage{txfonts}

\usepackage{natbib}
\bibpunct{(}{)}{;}{a}{}{,} 

\begin{document}

   \title{Coupling hydrodynamics with comoving frame radiative transfer}

   \subtitle{I. A unified approach for OB and WR stars}

   \author{A. A. C. Sander
            \and
           W.-R. Hamann
            \and
           H. Todt
            \and
           R. Hainich
            \and
           T. Shenar
          }

   \institute{Institut f\"ur Physik und Astronomie, Universit\"at Potsdam,
              Karl-Liebknecht-Str. 24/25, D-14476 Potsdam, Germany\\
              \email{ansander@astro.physik.uni-potsdam.de}
             }

        \date{Received February 17, 2017/ Accepted April 18, 2017}

\abstract{
  For more than two decades, stellar atmosphere codes have been used to derive the stellar and wind 
  parameters of massive stars. Although they have become a powerful tool and sufficiently
  reproduce the observed spectral appearance, they can hardly be used for more than measuring parameters.
  One major obstacle is their inconsistency between the calculated
  radiation field and the wind stratification due to the usage of prescribed mass-loss rates and
  wind-velocity fields. 
}{
  We present the concepts for a new generation of hydrodynamically
        consistent non-local thermodynamical 
        equilibrium (non-LTE) stellar atmosphere models that allow for
        detailed studies of radiation-driven stellar winds. As a first demonstration,
        this new kind of model is applied to a massive O star.
}{
  Based on earlier works, the PoWR code has been
  extended with the option to consistently solve the hydrodynamic
  equation together with the statistical equations and the radiative transfer 
        in order to obtain a hydrodynamically consistent atmosphere stratification.
  In these models, the whole velocity field is iteratively updated together
        with an adjustment of the mass-loss rate.
}{
  The concepts for obtaining hydrodynamically consistent models using a 
        comoving-frame radiative transfer are outlined. To provide a useful benchmark, 
        we present a demonstration model, which was motivated to describe the well-studied 
        O4 supergiant $\zeta\,$Pup. The obtained stellar
        and wind parameters are within the current range of literature values. 
}{
  For the first time, the PoWR code has been used to obtain a hydrodynamically
        consistent model for a massive O star. This has been achieved by a profound
        revision of earlier concepts used for Wolf-Rayet stars.
        The velocity field is shaped by various elements contributing to the radiative
        acceleration, especially in the outer wind. The results further indicate that for more 
        dense winds deviations from a standard $\beta$-law occur.
}


\maketitle


\section{Introduction}
  \label{sec:intro}

  In order to understand massive stars and their winds, stellar atmosphere models
        have become a powerful and widely used instrument. Typically applied for spectroscopic
        analysis, these models yield quantitative information on the stellar wind properties 
        together with fundamental stellar parameters. The special conditions in stellar 
        winds lead to the development of sophisticated model atmosphere codes performing
        several complex calculations. The outer layers are not even close to local thermodynamical 
        equilibrium (LTE), requiring the population numbers to be calculated from a set of 
        statistical equations. For a sufficient treatment, large model atoms with hundreds of levels 
        in total have to be considered. Furthermore, the line-driven winds of hot stars demand a 
        proper description of the radiative transfer in an expanding atmosphere.
        
        Basically three different approaches exist to tackle the radiative transfer problem:
        The first are analytical descriptions, usually based on the concept of CAK theory \citep[named after
        pioneering work of][]{CAK1975}, which allow for rapid calculation of the radiative acceleration
        at the cost of several approximations. While the theory has undergone several extensions relaxing 
        the original assumptions \citep[see, e.g.,][]{FA1986,Pauldrach+1986,Kudritzki+1989,
        Gayley1995,Puls+2000,Kudritzki2002}, it has opened up the whole area of time-dependent and
        even multi-dimensional calculations \citep[e.g.,][]{Owocki+1988,Feldmeier1995,OP1999,DO2005,
        Sundqvist+2010}. CAK-like concepts are therefore not only used in stellar atmosphere analyses, 
        but also in most cases where a more detailed radiative transfer would be too costly from a 
        computational standpoint; for example, detailed multi-dimensional hydrodynamical simulations of a stellar
        cluster or the interaction with a companion \citep[e.g.,][]{Blondin+1990,Manousakis+2012,vanMarle+2012,CH2015}. 
        
  An alternative to (semi-)analytical approximations is to calculate the radiative force with the
        help of Monte Carlo (MC) methods. Motivated already by the work of \citet{LS1970}, this 
        approach was first applied by \citet{AL1985} and later used for a variety of mass-loss
        studies \citep[e.g.,][]{deKoter+1993,deKoter+1997,Vink+1999,Vink+2000,Vink+2001}. More recently the
        application has been widely extended, including velocity field and clumping studies as well as
        multi-dimensional calculations \citep[e.g.,][]{MV2008,MV2014,Muijres+2011,Muijres+2012,Surlan+2012,Noebauer+2015}.
        Using the MC approach allows one to include effects such as multiple line scattering, but is
        computationally much more expensive than CAK-like calculations, especially in multi-dimensional 
        approaches. It is therefore mostly used for fundamental studies and rarely applied when analyzing a
        particular stellar spectrum.
        
        A third method to tackle the radiative transfer is the calculation in the comoving frame (CMF).
        Built on the conceptual work of \citet[][]{Mihalas+1975}, it is essentially a brute-force integration
        over the frequency range, using the advantage that opacity $\kappa$ and emissivity $\eta$ are
        isotropic in the comoving frame. Various studies using a CMF approach have been performed since
        the 1980s \citep[e.g.,][]{Hamann1980,Hamann1981,Hillier1987,Pauldrach+1986,Sellmaier+1993,Baron+1996},
        culminating in the development of a handful of stellar atmosphere codes using the CMF
        radiative transfer either partially or exclusively, including \textsc{phoenix} \citep{Hauschildt1992,HB1999},
        \textsc{wmbasic} \citep{Pauldrach+1994,Pauldrach+2001}, \textsc{fastwind} \citep{Santolaya-Rey+1997,Puls+2005},
        \textsc{cmfgen} \citep{Hillier1990a,Hillier1990b,HM1998}, and \textsc{PoWR} \citep{H1985,H1986,GKH2002,HG2003}. For studying stars
        with denser winds and especially Wolf-Rayet stars, almost all spectral analyses are performed
        with CMF-based atmosphere codes \citep[e.g.,][]{HM1999,Crowther+2002,HGL2006,SHT2012,Hainich+2014}.
        Due to the complexity of the CMF calculation, these codes  typically assume spherical
        symmetry, allowing for a $1$- or $1.5$-dimensional treatment, and a stationary wind situation.

        Given that on top of the two major tasks, that is, solving the statistical equations and the radiative transfer, 
        several further challenges exist, such as       iron-line blanketing or the need for a consistent calculation
        of the temperature stratification in an expanding, non-LTE environment, it is not that surprising that
        only a few codes exist that can adequately model a stellar atmosphere for a hot star with a dense, 
        line-driven wind. So far, these codes 
        are typically used for either measuring stellar and wind parameters or predicting fluxes and related 
        quantities for a given set of parameters. Yet, only a few examples exist where they are used 
        to actually predict the wind parameters. This lack of examples results from
        the fact that -- at least in the wind part -- most stellar atmosphere models use a prescribed
        velocity field instead of consistently calculating the wind stratification. While this approach is mostly 
        sufficient for the current use of the atmosphere
        models, it also cuts off a variety of potential applications. In order to open up this
        perspective, we present a new approach for hydrodynamically consistent stellar atmosphere models
        using the Potsdam Wolf-Rayet (PoWR) model atmosphere code. Originally starting from earlier efforts
        for Wolf-Rayet (WR) stars, we have developed a brand new scheme to update the mass-loss rate and
        the velocity stratification that can finally be applied to both WR and OB models. For the first time,
        we will present a hydrodynamically consistent PoWR model for an O supergiant, closely reproducing 
        most of the spectral features for $\zeta$\,Pup.
        
In Sect.\,\ref{sec:hydro} of this work, we briefly discuss the underlying stationary
wind hydrodynamics and introduce a special notation that is helpful in analyzing the status 
of PoWR models with respect to hydrodynamical consistency. A special emphasis is given in
the following Sect.\,\ref{sec:critpoint} on the meaning of the critical point. Afterwards, the
basic concepts of the PoWR code and its set of input parameters are outlined in Sect.\,\ref{sec:powr}.
Sect.\,\ref{sec:hdmodels} then deals with all the techniques used for obtaining
a hydrodynamically consistent model before showing and discussing the results for an
example model in Sect.\,\ref{sec:results}. Finally, conclusions are drawn in Sect.\,\ref{sec:conclusions}.

\section{Stationary wind hydrodynamics}
  \label{sec:hydro}

In a hot stellar wind, which we here describe as a one-dimensional, stationary outflow, the accelerations due to radiation and 
gas pressure have to balance gravity $g(r) = GM_\ast r^{-2}$ and inertia $\varv \frac{\mathrm{d} \varv}{\mathrm{d} r}$. 
The corresponding equation of motion is therefore
\begin{equation}
  \label{eq:simplehydro}
  \varv \frac{\mathrm{d} \varv}{\mathrm{d} r} + \frac{GM}{r^2} = a_\text{rad}(r) + a_\text{press}(r)
,\end{equation}
with $a_\text{rad}$ representing the total radiative acceleration, that is,
\begin{align}
  \label{eq:aradparts}
  a_\text{rad}(r) := &~ a_\text{lines}(r) + a_\text{cont}(r) \\
                               = &~ a_\text{lines}(r) + a_\text{true cont}(r) + a_\text{thom}(r)\text{,}
\end{align}
using the same notation as in \citet{Sander+2015}. The term  $a_\text{press}$ describes 
the gas (and potentially turbulence) pressure, that is,
\begin{equation}
  \label{eq:apressdef}
     a_\text{press}(r) := -\frac{1}{\rho} \frac{\mathrm{d}P}{\mathrm{d}r}.
\end{equation} 
To replace the pressure $P$ with the density, we use the equation of state for an ideal 
gas $P(r) = \rho(r) \cdot a^2(r)$ and define 
  \begin{equation}
    a^2(r) := \frac{k_\text{B} T(r)}{\mu(r)\,m_\text{H}} + \frac{1}{2} \varv_\text{mic}^2
  ,\end{equation}
with $\mu$ being the mean particle mass (including electrons)
in units of the hydrogen atom mass $m_\text{H}$, $T$ being the electron temperature, 
and $\varv_\text{mic}$ the microturbulence velocity, which is a free parameter in our models. 
For vanishing turbulence, $a(r)$ is identical to the isothermal sound speed.

With the equation of state together with the equation of continuity
   \begin{equation}
    \label{eq:cont}
    \dot{M} = 4 \pi r^2 \varv(r)\,\rho(r) \text{,}
  \end{equation}
the $a_\text{press}$-term in Eq.\,(\ref{eq:simplehydro}) can be rewritten in three terms
that remove all explicit $\rho$-dependencies in favor of terms containing only $\varv$, $r$,
and $a$ \citep[see, e.g.,][for an explicit calculation]{Sander+2015}.
As a consequence, the hydrodynamic equation for a spherically-symmetric wind reads
  \begin{align}
          \label{eq:stdhydro}
     \varv \left( 1 - \frac{a^2}{\varv^2} \right) \frac{\mathrm{d} \varv}{\mathrm{d} r} & = a_\text{rad} - g + 2 \frac{a^2}{r} - \frac{\mathrm{d} a^2}{\mathrm{d} r} \\
          \label{eq:stdhydroG}
                                                                        & = \frac{GM}{r^2} \left(\Gamma_\text{rad} - 1\right) + 2 \frac{a^2}{r} - \frac{\mathrm{d} a^2}{\mathrm{d} r}
  \end{align}
with $\Gamma_\text{rad}(r) := a_\text{rad}(r) / g(r)$. For an easier reading, we have dropped the explicit notation of the radius dependencies in Eqs.\,(\ref{eq:stdhydro}) and (\ref{eq:stdhydroG}) and continue to do so unless absolutely necessary for the context.
 Using the two definitions
\begin{align}
  \label{def:ftilde}
        \mathcal{\tilde{F}} & := 1 - \Gamma_\text{rad} - 2 \frac{a^2 r}{GM} + \frac{r^2}{GM} \frac{\mathrm{d} a^2}{\mathrm{d} r}  &  \text{and} \\
        \label{def:gtilde}
        \mathcal{\tilde{G}} & := 1 - \frac{a^2}{\varv^2}  \text{,} &
\end{align}
one can write Eq.\,(\ref{eq:stdhydro}) in a more compact way:   
  \begin{align}
     r^2 \varv \left( 1 - \frac{a^2}{\varv^2} \right) \frac{\mathrm{d} \varv}{\mathrm{d} r} & =  GM \left(\Gamma_\text{rad} - 1 \right) + 2 a^2 r - r^2 \frac{\mathrm{d} a^2}{\mathrm{d} r} \\
                 r^2 \varv~\mathcal{\tilde{G}} \frac{\mathrm{d} \varv}{\mathrm{d} r} & = - GM~\mathcal{\tilde{F}} \\
                \label{eq:hdfg}
                  \frac{\mathrm{d} \varv}{\mathrm{d} r} & = - \frac{g}{\varv} \frac{\mathcal{\tilde{F}}}{\mathcal{\tilde{G}}}.
  \end{align}
        
        The quantities $\mathcal{\tilde{F}}$ and $\mathcal{\tilde{G}}$ are dimensionless and therefore ideal for visualizations.
        We note that in the subsonic regime, that is, $\varv \ll a$, we obtain $\mathcal{\tilde{G}} \rightarrow -a^2/\varv^2$ and thus Eq.\,(\ref{eq:hdfg})
        reduces to 
        \begin{equation}
          \label{eq:hystlimit}
                  \frac{\mathrm{d} \varv}{\mathrm{d} r} = \frac{g\,\varv}{a^2}~\mathcal{\tilde{F}} \text{,}
        \end{equation}
        which is a form of the hydrostatic equation discussed in a previous paper \citep{Sander+2015}.

\section{The critical point}
  \label{sec:critpoint}
        
        The crucial difference between Eq.\,(\ref{eq:hystlimit}) and the full hydrodynamic equation (\ref{eq:hdfg})
        is the denominator $\mathcal{\tilde{G}}$. In contrast to the quasi-hydrostatic case, the hydrodynamic equation has a critical
        point at $\varv = a$, that is, where the denominator $\mathcal{\tilde{G}}$ becomes zero. In order to allow for a finite
        solution for the velocity gradient at the corresponding radius $r_\text{c}$, the nominator $\mathcal{\tilde{F}}$ must
        also vanish at exactly this point. This leads to the constraint 
        \begin{equation}
          \label{eq:critpoint}
        \mathcal{\tilde{F}}(r_\text{c}) \stackrel{!}{=} \mathcal{\tilde{G}}(r_\text{c}) \stackrel{!}{=} 0 \text{.}
        \end{equation}
        
        As will be discussed below, this constraint implies a fixing of the mass-loss rate $\dot{M}$,
        even though this quantity does not appear explicitly in the hydrodynamic Eq.\,(\ref{eq:stdhydro}). Since
        the hydrostatic equation does not have this constraint, it can provide a solution for $\varv(r)$ for any 
        non-vanishing value of $\dot{M}$.
                
\begin{figure}[ht]
  \resizebox{\hsize}{!}{\includegraphics{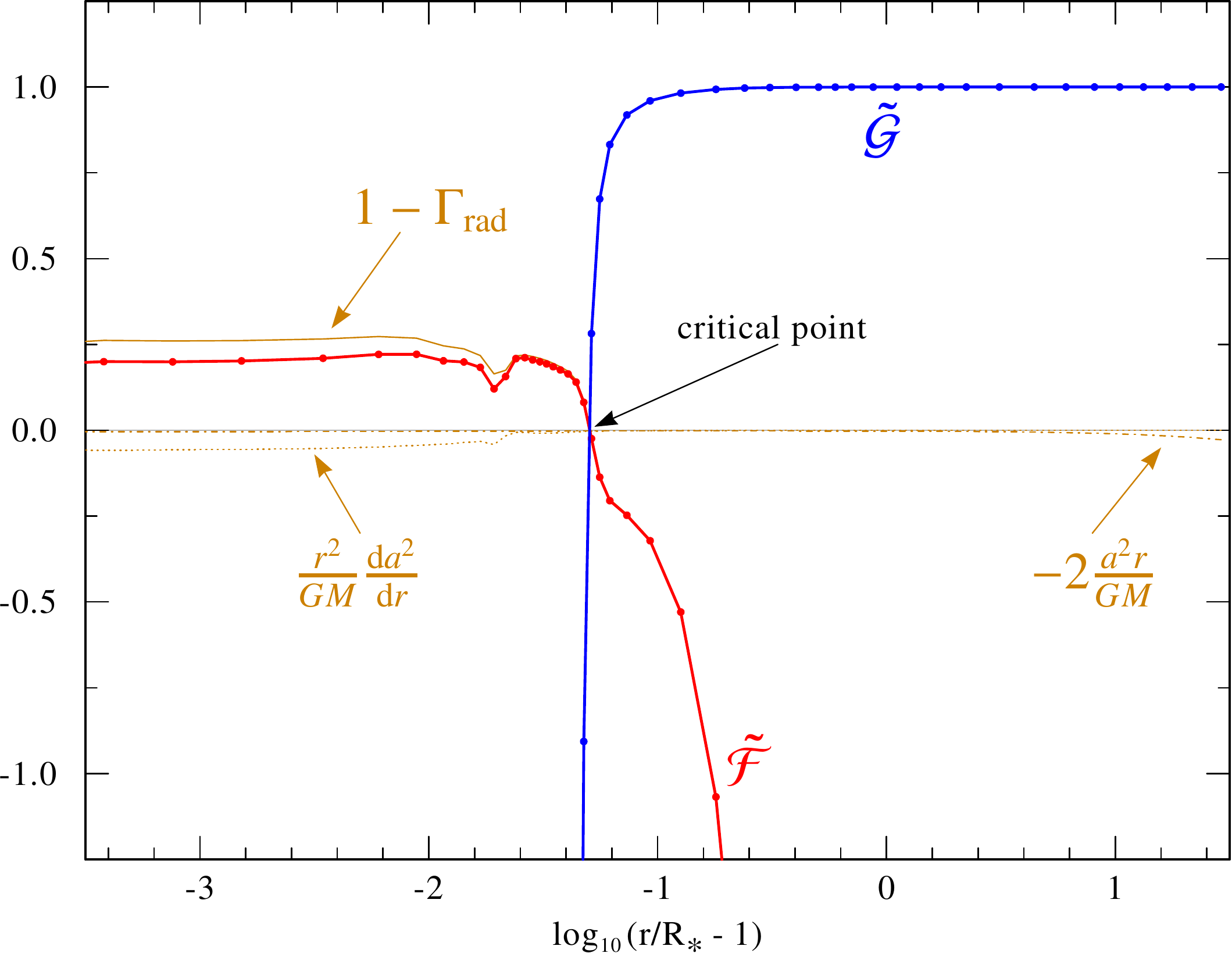}}
  \caption{The dimensionless depth-dependent quantities $\mathcal{\tilde{F}}$ and $\mathcal{\tilde{G}}$
                  are shown for a converged hydrodynamically consistent atmosphere model. The three terms adding up 
                                         to $\mathcal{\tilde{F}}$ are also indicated by the orange curves. At the critical point,
                                         both $\mathcal{\tilde{F}}$ and $\mathcal{\tilde{G}}$ become zero.}
  \label{fig:zpupfg}
\end{figure}
        
        An illustration of $\mathcal{\tilde{F}}(r)$ and $\mathcal{\tilde{G}}(r)$ for a converged hydrodynamically consistent
        model is shown in Fig.\,\ref{fig:zpupfg}.
  Unlike in the CAK-type approaches, the critical point in Eq.\,(\ref{eq:hdfg}) and Fig.\,\ref{fig:zpupfg} is 
        identical to the sonic point, potentially corrected for a turbulence contribution.
        This is a direct consequence of our approach where we do not assume any semi-analytical expression for the radiative 
        acceleration $a_\text{rad}$. Instead, we simply treat $a_\text{rad}(r)$ as a given quantity that can
        be expressed as a function of radius. Of course $a_\text{rad}$ 
        also reacts on changes of the velocity field and the mass-loss rate, but instead of trying to parametrize
        this in analytical form, we use an iterative approach and recalculate $a_\text{rad}(r)$ after any
        adjustment of $\varv(r)$ or $\dot{M}$ until the thereby calculated acceleration does not enforce further
  velocity or mass-loss rate updates. 
        
        The PoWR code has already been used in the past for hydrodynamically consistent calculations of WR stars by \citet{GH2005,GH2008}.
        In their approach, which turned out to be suitable only for thick-wind WR stars, the critical point was not 
        identical to the sonic point due to a semi-analytical approach where the radiative acceleration was described with the
        help of an effective force multiplier parameter $\alpha(r)$. In the present work, we do not use such a parametrization
        and therefore the critical point in our hydrodynamic equation is identical to the sonic point. The 
        implementation of the new method is also conceptually different from the one described in \citet{GH2005} and
        will be further outlined in Sect.\,\ref{sec:hdmodels}.

\section{PoWR}
  \label{sec:powr}

  \subsection{Basic concepts}
    \label{sec:basics}

  For the Potsdam Wolf-Rayet (PoWR) model atmospheres we assume a spherically symmetric atmosphere 
        with a stationary mass outflow. In order to properly describe the situation of an expanding 
        atmosphere without the LTE approximation, the 
        equations of statistical equilibrium and radiative transfer have to be solved iteratively until a consistent 
        solution for the radiation field and the population numbers are obtained. In addition, the 
        temperature stratification is updated iteratively to ensure energy conservation in the expanding 
        atmosphere. This is performed using the improved Uns{\"o}ld-Lucy method described in \citet{HG2003} 
        or alternatively via the electron thermal balance which has recently been added to the 
        PoWR code \citep[see][and references therein]{Sander+2015}.
        
        If then all changes to the population numbers are smaller than a defined threshold, the atmosphere model is
        considered to be converged and the synthetic spectrum is calculated using a formal integration in the observer's frame.
        The iron group elements are treated in a superlevel approach: The levels are
        grouped into energy bands, which are then represented by superlevels. While we assume LTE for the relative occupations
        of the individual levels inside a superlevel, the superlevels themselves are treated in full non-LTE. The detailed
        cross-sections for the superlevel transtions have been prepared on a sufficiently fine frequency grid prior to the model iteration and 
				contain all the individual transitions to ensure that radiative transfer treats each of these transitions at their proper frequency. 
				\citep[See][for more details]{GKH2002}.
        PoWR is furthermore able to account for wind inhomogeneities in the so-called ``microclumping'' approximation \citep[see][]{HK1998}. 
				In the calculation of the formal integral, PoWR can also account for optically thick clumps in an approximate way, 
        see \citet{OHF2007} for details.
        
        The radiative transfer is calculated in the CMF, thereby implicitly accounting for
        multiple scattering and avoiding all simplifications, which are done in the faster but more approximate
        concepts used, for example, in time-dependent calculations. In particular, the solution is obtained by solving the moment 
        equations via a differencing scheme based on the concepts of \citet{Mihalas+1976b,Mihalas+1976c}. The 
        variable Eddington factors required in this scheme are obtained from a so-called ``ray-by-ray solution'' where 
        an angle-dependent radiation transfer is performed using short-characteristics integration \citep[see][for details]{KHG2002}.
        The CMF radiative transfer requires a strictly monotonic velocity field. Even in a
        stationary wind, this might not always be the case and therefore in some cases the solution for $\varv(r)$ 
        resulting from the hydrodynamic equation cannot be applied. This will be discussed in more detail in
        a future paper and does not apply to the models presented in this work.
        
        In the case of stars with low or moderate $\log g$ it is sufficient to assume that the intrinsic line profiles
  are Gaussians with a constant Doppler broadening velocity $\varv_\text{dop}$ during the CMF calculations
        while in the formal integral, where the emergent spectrum is eventually obtained, detailed thermal, 
        microturbulent, and pressure broadening are accounted for with their depth dependence. 
        
        The necessary atomic data required for our calculations are taken from a variety of sources, combining 
  \citet{Wiese+1966book}, \citet{Eissner+1974}, \citet{BS1975book}, the opacity project \citep[OP, ][]{CM1992TOPBASE}, the
        Kurucz atomic database\footnote{\texttt{http://kurucz.harvard.edu/atoms.html}}, the NIST atomic
        database\footnote{\texttt{https://www.nist.gov/pml/atomic-spectra-database}}, private communication with
        K.~Butler, and several minor sources listed in \citet{Hamann+1992}. For argon, which turns out to be one of the
        more important elements driving the outer wind of our demonstration model, we use opacity project data combined with
        level energies from the NIST atomic database. The iron group elements, which are 
        treated as one generic element with the help of the superlevel approach described in detail in \citet{GKH2002},
        are modeled using Kurucz data if available (usually up to ionization stage X), 
        while opacity project data are applied to also cover the higher ions.
        The collisional cross-sections are described with different formulae depending on the element and ion,
        most notably from \citet[Eqs.\,6.24, 6.25]{Jefferies+1968}, K.~Butler (priv. comm.), and \citet[Eq.\,22]{vR1962}. The latter is also used
        for all collisional transitions of argon and the iron group superlevels if the corresponding radiative transition is allowed.
        The collisional cross-sections of forbidden transitions are mostly approximated by \citet[appendix 4]{Mendoza1983}, though for very few ions
        a more specialized treatment is applied \citep[e.g., ][for \ion{He}{i}]{BFK1982}. For the bound-free transitions, we
        use \citet[Eq.\,6.39]{Jefferies+1968} for all collisional ionizations while we branch between OP data fits, \citet{Mihalas1967book} and
        \citet{Seaton1960}, depending on the ion for the photoionization cross-sections. The hydrogenic 
        approximation \citep[e.g.,][]{Cowan1981book} is widely used as fallback, which also applies for Ar and the iron
        group.

  \subsection{Model parameters}
          \label{sec:powrparams}
                
                PoWR model atmospheres can be specified by a set of fundamental parameters. These are:
        
                \begin{itemize}
      \item The chemical abundances of all considered elements, typically given as mass fractions $X_i.$ 
                        \item Two out of the three quantities connected by Stefan-Boltzmann's law, namely:
                                \begin{itemize}
                                        \item The stellar radius $R_\ast$, defined at a specified Rosseland continuum optical depth $\tau_\ast$ (default: $\tau_\ast = 20$). 
                                        \item The effective temperature $T_\ast$ at the radius $R_\ast.$
                                        \item The luminosity $L_\ast = 4 \pi R_\ast^2 \sigma_\textsc{sb} T_\ast^4$.
                                \end{itemize}      
      \item The stellar mass $M_\ast$, either given explicitly via input of $M_\ast$ or $\log g$, or 
            calculated from the luminosity if the stellar mass is not given otherwise. In the latter case, depending on the stellar type, the 
						mass-luminosity-relations from \citet{L1989} or \citet{Graefener+2011} are used.
      \item The mass-loss rate $\dot{M}$ or an implying quantity (see below).
      \item The terminal wind velocity $\varv_\infty$ and the wind velocity law $\varv(r)$, directly implying the density stratification 
                              via the continuity Eq.\,(\ref{eq:cont}).
                        \item The clump density contrast $D(r) = \rho_\text{cl}(r)/\overline{\rho}(r) = f_\text{V}^{-1}(r)$ \citep[cf.][for a detailed description]{HK1998}.
    \end{itemize}
  
        As an alternative to the mass-loss rate $\dot{M}$, one can also specify a line emission
  measure in the form of either the \emph{transformed radius}
  \begin{equation}
    \label{def:rt}
    R_\text{t} := R_\ast \left[ \frac{\varv_\infty}{2500\,\text{km/s}} \left/ \frac{\dot{M}\sqrt{D}}{10^{-4}\,\text{M}_\odot\text{/yr} } \right. \right]^{\frac{2}{3}}
  ,\end{equation}
  \citep[][for the current form]{SHW1989,HK1998} or the \emph{wind strength parameter}
  \begin{equation}
          \label{def:qws}
     Q_\text{ws} := \frac{\dot{M}\sqrt{D}}{\left(R_\ast \varv_\infty\right)^{3/2}}
  ,\end{equation}
  \citep[][for the current form]{Puls+1996,PVN2008}. Since all other quantities in Eqs.\,(\ref{def:rt}) and (\ref{def:qws}) 
        have to be specified anyhow, these quantities imply a certain value of $\dot{M}$. Using $R_\text{t}$ or $\log Q_\text{ws}$ can
  be helpful when calculating model grids or searching for models with a similar emission line strength in their normalized spectra.

\begin{figure}[ht]
  \resizebox{\hsize}{!}{\includegraphics{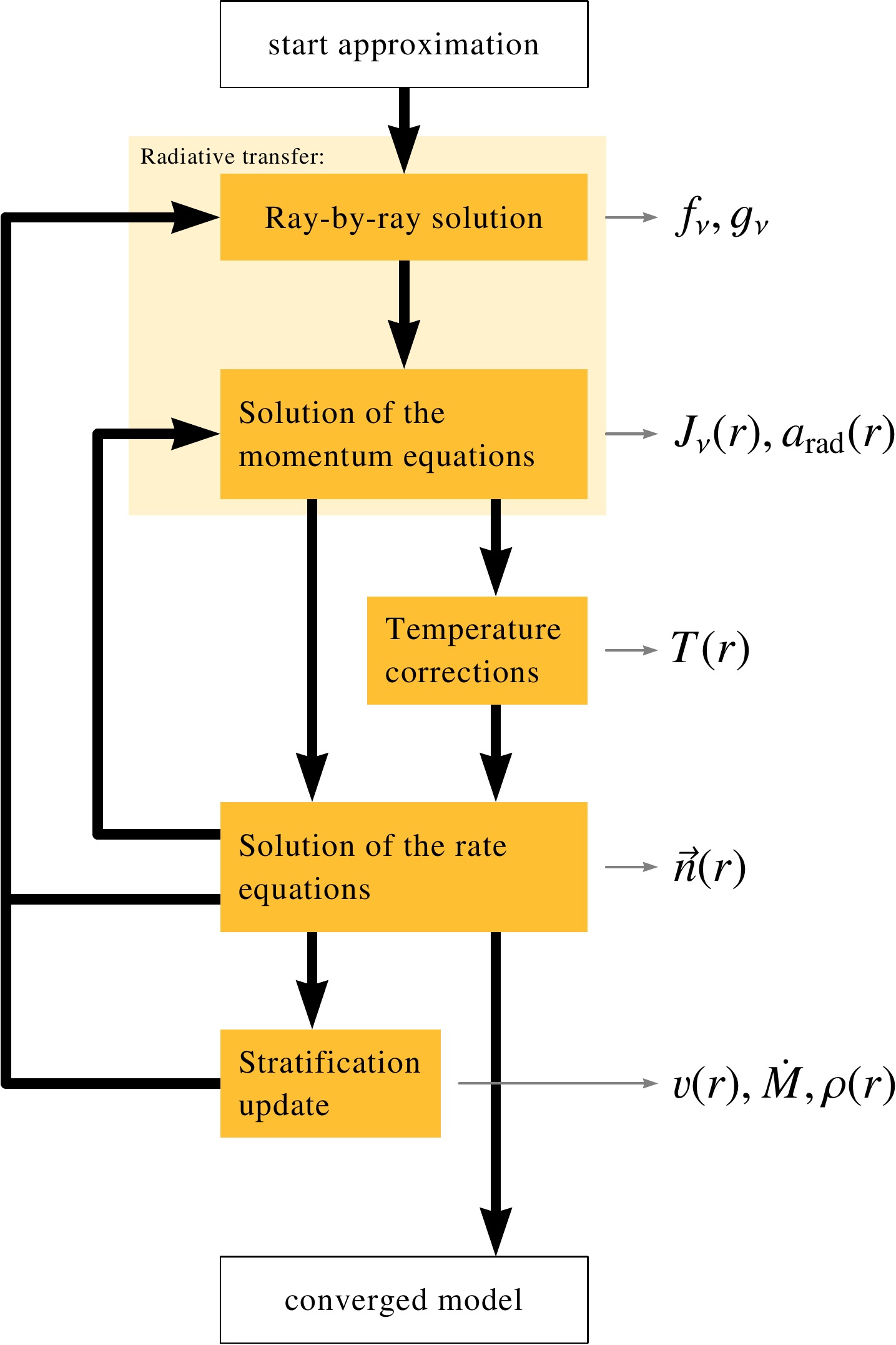}}
  \caption{Iteration scheme for the calculation of a PoWR atmosphere model. The inner cycle without the ray-by-ray radiative transfer
                 is typically applied for a few (typically $5$-$6$) iterations before the Eddington factors must be renewed. If the criteria for a 
                                         stratification update are fulfilled -- see Sect.\,\ref{sec:hdscheme} for the case of a HD update -- this is performed before
                                         the next ray-by-ray transfer would be scheduled. On the right side of the figure, indicated by gray arrows, 
                                         the most important quantities obtained by the current step are highlighted.}
  \label{fig:flowchart}
\end{figure}

  In hydrodynamically consistent models, $\dot{M}$ and $\varv(r)$ are adjusted in order to
  ensure that the hydrodynamic equation is fulfilled throughout the atmosphere. However, as they also
  define the density stratification, starting values are still required as an input for the calculations. Depending on
        the starting model, the resulting $\dot{M}$ and $\varv(r)$ of a converged hydrodynamically consistent model
        can differ significantly from their initial specifications.

  \subsection{Iteration scheme}
    \label{sec:itscheme}
  With the main concepts and parameters given in the previous paragraphs, the overall iteration scheme for a PoWR
        model can now be summarized by the following steps:
        
  \begin{enumerate}
    \item Model start: Setup of radius and frequency grids, first velocity stratification, start approximation for 
                      the population numbers $\vec{n}(r)$ and the radiation field $J_\nu$(r)
    \item Main iteration
      \begin{enumerate}
        \item Solution of the radiative transfer in the comoving frame
        \item Temperature corrections (if necessary)
        \item Solution of the statistical equations
        \item (optional:) Solution for the hydrostatic or hydrodynamic equation to update the velocity/density stratification 
      \end{enumerate} 
    \item Formal integration: Calculation of the emergent spectrum in the observer's frame
  \end{enumerate}
 
   For efficiency, we use the method of variable Eddington factors, where 
         the Eddington factors $f_\nu$ and $g_\nu$, which have to be obtained from the more costly ray-by-ray
         radiative transfer, are only updated every few iterations while otherwise only the momentum equations
         are solved in order to obtain the new radiation field and the radiative acceleration. 
         For convenience, we schedule stratification updates immediately before the next renewal of the Eddington factors.
        
   A sketch of this iteration scheme is given in Fig.\,\ref{fig:flowchart}. The stratification
         update, which can be either restricted to the quasi-hydrostatic part as outlined in \citet{Sander+2015}
         or the full hydrodynamical update described in this work, is fully integrated into the main iteration. 
         The particular details of the hydrodynamic stratification update
         are discussed in Sect.\,\ref{sec:hdmodels}.

\section{Hydrodynamically consistent models}
  \label{sec:hdmodels}

  \subsection{Start approximation}
          \label{sec:hdstart}

  In order to integrate the hydrodynamic equation in the form of Eq.\,(\ref{eq:hdfg}), the
        quantities $a(r)$ and $a_\text{rad}(r)$ have to be specified as a function of radius. This
        cannot be done from scratch and therefore a starting approximation for the stellar atmosphere
        has to be given, including a velocity stratification. Unless the new model is only a small
        variation of an already existing hydrodynamically consistent model, where one could employ
        the old velocity field, usually a model with a $\beta$-law connected to a consistent 
        hydrostatic solution \citep[see]{Sander+2015} is adopted as a starting approximation.
        For the mass-loss rate, it has turned out to be
        helpful, if at least the global energy budget is close to consistency. To obtain this 
        budget, the hydrodynamic equation is written in the form
\begin{align}
  \label{eq:hdbalance} 
  \varv \frac{\mathrm{d} \varv}{\mathrm{d} r} + \frac{G M_{\ast}}{r^2}  & =  a_\text{rad} - \frac{1}{\rho} \frac{\mathrm{d} P}{\mathrm{d} r},
\end{align}
and then integrated over $r$ and multiplied with $\dot{M}$:
\begin{align}
  \dot{M} \int \left( \varv \frac{\mathrm{d} \varv}{\mathrm{d} r} + \frac{G M_{\ast}}{r^2} \right) \mathrm{d} r & = \dot{M} \int \left(  a_\text{rad} - \frac{1}{\rho} \frac{\mathrm{d} P}{\mathrm{d} r} \right) \mathrm{d} r \nonumber \\
  \label{eq:workbalance}
  L_\text{wind} & = W_\text{wind}.
\end{align}
This last equation (Eq.\ref{eq:workbalance}) describes the balance between the modeled wind luminosity $L_\text{wind}$
and the provided power $W_\text{wind}$. Dividing Eq.\,(\ref{eq:workbalance}) by $L_\text{wind}$
yields the so-called \emph{work ratio}
\begin{equation}
  Q := \frac{W_\text{wind}}{L_\text{wind}}\text{.}
\end{equation}
While stellar atmosphere models with $Q < 1$ do not provide a radiative acceleration that is sufficient
to drive the wind, models with $Q > 1$ exhibit a radiation acceleration that could actually drive a stronger 
wind. Models with $Q = 1$ exactly supply the power that is required to drive the wind. The corresponding
models are therefore consistent on a ``global'' scale, as they fulfill the integrated form of the hydrodynamic 
equation. Although they usually do not fulfill this equation locally, models with $Q \approx 1$ are usually 
well suited as a starting model for the full hydrodynamic calculations.
A similar approach to identify proper start approximations has been used by \citet{GH2005} in their 
calculation of a hydrodynamically consistent WC atmosphere.

  \subsection{Obtaining a consistent velocity field}
        
        With a given starting model, all terms in the hydrodynamic equation are known, including the
        radiative acceleration $a_\text{rad}(r)$ as a function of radius, and the terms $\mathcal{\tilde{F}}(r)$
        and $\mathcal{\tilde{G}}(r)$ can be calculated. Since the ratio of the latter two essentially defines
        the right hand side of Eq.\,(\ref{eq:hdfg}), both terms have to vanish at exactly the same radius $r_\text{c}$,
        defining the critical point of the equation. For the non-consistent starting model, this is generally
        not the case and thus the radius $r_\mathcal{\tilde{G}} := r(\mathcal{\tilde{G}} = 0)$ will differ
        from the radius $r_\mathcal{\tilde{F}} := r(\mathcal{\tilde{F}} = 0)$. In some cases, $\mathcal{\tilde{F}}$
        can become zero at more than one point, thus indicating a non-monotonic solution for $\varv(r)$. While this does
        not prevent the integration of the hydrodynamic equation, a non-monotonic $\varv(r)$ cannot be used in 
        the CMF radiative transfer and thus such cases are discarded at the moment.

\begin{figure}[ht]
  \resizebox{\hsize}{!}{\includegraphics{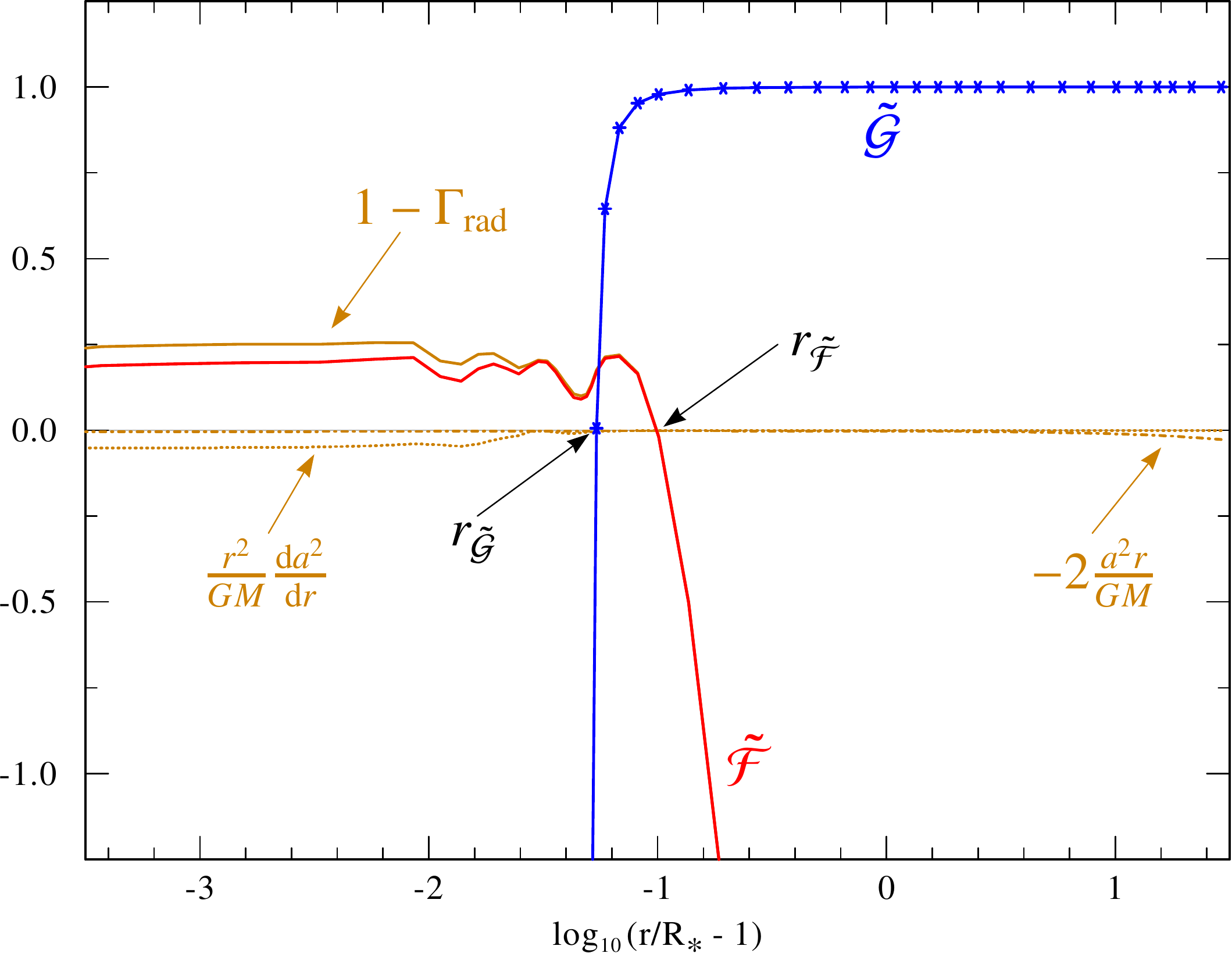}}
  \caption{As in Fig.\,\ref{fig:zpupfg}, but now for a standard PoWR model where only the quasi-hydrostatic
                       regime is treated self-consistently. The dimensionless, depth-dependent quantities 
                                         $\mathcal{\tilde{F}}$ and $\mathcal{\tilde{G}}$ now cross zero at different radii $r_\mathcal{\tilde{F}}$ 
                                         and $r_\mathcal{\tilde{G}}$ , which is typical
                                    for non-consistent models and indicates a necessary adjustment of the mass-loss rate.}
  \label{fig:fgnonhd}
\end{figure}

  Assuming that there is only one radius $r_\mathcal{\tilde{F}}$ at which we have $\mathcal{\tilde{F}} = 0$, 
        this point acts as the current candidate for the critical point. Starting at $r_\mathcal{\tilde{F}}$ with
        $\varv(r_\mathcal{\tilde{F}}) = a(r_\mathcal{\tilde{F}})$, Eq.\,(\ref{eq:hdfg})
        is then integrated inwards and outwards to obtain the new velocity field. By setting the wind velocity to
        the current value of $a$ at $r_\mathcal{\tilde{F}}$, the critical point condition is automatically fulfilled and we achieve a velocity
        field that smoothly passes through the critical point. In principle it would also be
        possible to start the integration at $r_\mathcal{\tilde{G}}$ or any point in-between $r_\mathcal{\tilde{F}}$
        and $r_\mathcal{\tilde{G}}$, but this would require a modification of $r_\mathcal{\tilde{F}}$ during the hydrodynamic iteration 
        in order to fulfill the critical point condition. Several approaches have been tested during the development phase
        and none of them turned out to be favorable. Essentially, the approach by \citet{GH2005,GH2008} made use
        of modifying $\mathcal{\tilde{F}}$ when changing the mass-loss rate. While this worked fine for some WR models, it
        turned out to fail for OB models. Furthermore, their approach required the calculation of a force multiplier parameter $\alpha(r)$, 
        which can only be obtained by a modified radiative transfer calculation, thereby essentially doubling the
        calculation times for the radiative transfer before each hydro stratification update. On the other hand,
  modifications of the $\Gamma_\text{rad}$-term in $\mathcal{\tilde{F}}$ without any prediction on how the radiative
        acceleration will react on changes of $\varv(r)$ or $\dot{M}$ turn out not to be precise enough. Thus the
        direct integration from $r_\mathcal{\tilde{F}}$ proved to be the best method, both in terms of
        stability and performance.

 \subsection{Calculation of the mass-loss rate}

  By starting the integration of the velocity field outwards from the critical point of the hydrodynamic equation,
        one automatically obtains the terminal velocity $\varv_\infty$ when reaching the outer boundary, since 
        $\varv_\infty \approx \varv(R_\text{max})$ as long as $R_\text{max}$ is chosen to be sufficiently large. This
        method so far provides a new velocity field that fulfills the hydrodynamic equation, but does not perform 
        any update of the mass-loss rate $\dot{M}$. This is already a powerful
        tool, but the major issue with such a solution is their inconsistency with some of the initial stellar
        parameters; since the integration starts
        from the critical point, also the inner boundary value $\varv_\text{min} := \varv(R_\ast)$ is not fixed, but
        instead obtained from integrating the hydrodynamic equation. As long as $r_\mathcal{\tilde{F}}$
        and $r_\mathcal{\tilde{G}}$ are not identical before the integration, the total optical depth at $R_\ast$ will
        change after the velocity update. This especially means that $T_\ast$ and the corresponding $R_\ast$ in a converged
        model would refer to a different optical depth than in the starting model. As long as the total
        optical depth is larger than the old one, one could infer the values for the original optical depth. However, the consequence 
        would be that the obtained hydrodynamic model refers to a different temperature and radius than the
        starting model, thereby being inconsistent with the radiative transfer calculation.

  This problem can be solved by introducing another constraint, namely the conservation of the total optical depth. More precisely,
        for practical reasons we demand the conservation of the total Rosseland continuum optical depth, which we denote as $\tau_\text{Ross}(R_\ast)$ 
        throughout this work. As a consequence, the mass-loss rate $\dot{M}$ needs to be updated, but
        the main model parameters $T_\ast$ and $R_\ast$ now keep their intended reference. Unfortunately, finding a proper update 
        method for $\dot{M}$ is not a trivial task. A simple approach would be to use the definition of the optical depth and replace 
        the density via
        Eq.\,\ref{eq:cont} to obtain an expression that explicitly contains $\dot{M}$:
  \begin{align}
    \tau_\text{Ross}(R_\ast) & = \int\limits_{R_\text{max}}^{R_\ast} \kappa_\text{Ross}(r)\,\mathrm{d} r \\
                             & = \int\limits_{R_\text{max}}^{R_\ast} \rho(r)~ \varkappa_\text{Ross}(r)\,\mathrm{d} r \\
   \label{eq:taumaxmdot}                                                 
                             & =  \frac{\dot{M}}{4\pi}\int\limits_{R_\text{max}}^{R_\ast} \frac{\varkappa_\text{Ross}(r)}{r^2~\varv(r)} \mathrm{d} r\text{.} 
  \end{align} 
  However, even though Eq.\,(\ref{eq:taumaxmdot}) seems like a straight-forward approach to extract the density and thus the mass-loss rate, one must keep in mind that $\varkappa_\text{Ross}(r)$ is not generally depth-independent, as some of the contributions (e.g., the bound-free opacities) do
        not just have a linear dependence on $\rho(r)$.
        In fact, using this expression leads to large changes in $\dot{M}$, often over-predicting the required changes by orders of magnitude.
        Since the critical point tends to change significantly even for moderate $\dot{M}$ updates, this method can only be successfully applied in very few
        cases and thus cannot be considered as a standard approach. The sensitivity of the critical point was already found by \citet{Pauldrach+1986} when 
        they used an early form of this concept with the Thomson opacity instead of the Rosseland continuum opacity as they 
        did not account for the free-free and bound-free continuum in their radiative force. Similar problems occur when employing other quantities
        with analog descriptions, such as the integrated density without the mass absorption coefficient
        \begin{equation}
    \label{eq:smax}
    s_\text{max} = \int\limits_{R_\text{max}}^{R_\ast} \rho(r)~\mathrm{d} r = \frac{\dot{M}}{4\pi} \int\limits_{R_\text{max}}^{R_\ast} \frac{1}{r^2~\varv(r)} \mathrm{d} r
  ,\end{equation}
        or the total atmosphere mass 
        \begin{equation}
  M_\text{atm} = 4\pi \int\limits_{R_\ast}^{R_\text{max}} \rho(r) r^2 \mathrm{d} r = \dot{M} \int\limits_{R_\ast}^{R_\text{max}} \frac{1}{\varv(r)} \mathrm{d} r\text{.}
\end{equation}

\begin{figure}[ht]
  \resizebox{\hsize}{!}{\includegraphics{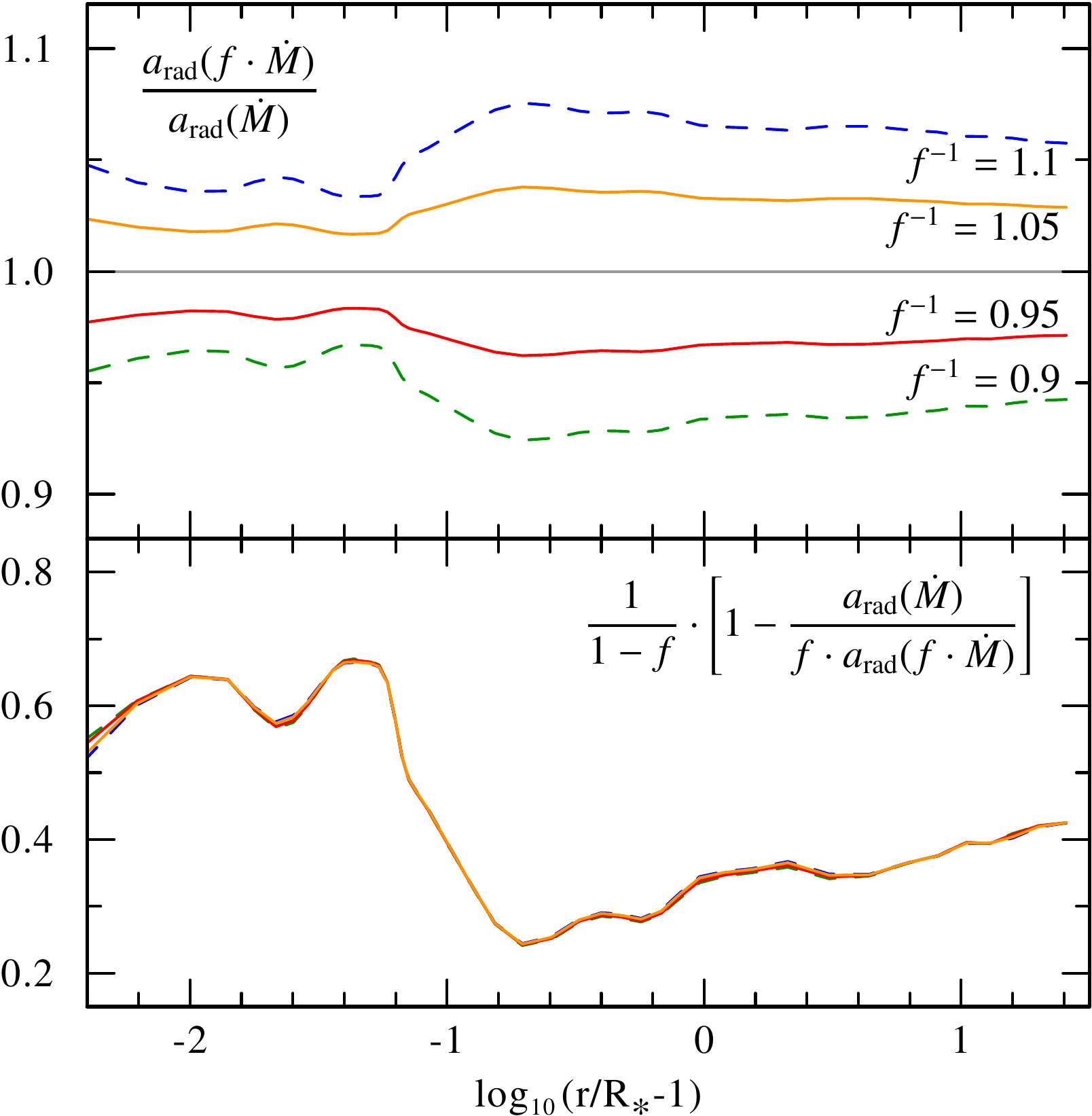}}
  \caption{Upper panel: The response of the radiative acceleration $a_\text{rad}(r)$ to a change of the mass-loss rate $\dot{M}$
                 is illustrated by plotting the ratio of the modified to unmodified acceleration for different modification factors $f$.
                                         Lower panel: The results for different factors $f$ can be scaled to an almost unique curve.}
  \label{fig:aradresponse}
\end{figure}

  In order to obtain a more stable method, this work utilizes a completely different approach, where
        we approximate the response of the radiative acceleration $a_\text{rad}$ to a change of the mass-loss rate by a factor $f$. A typical
        example for an O-star model is shown in Fig.\,\ref{fig:aradresponse}, where the ratio of unmodified to modified acceleration
        is shown for different values of $f^{-1}$. As the radiative acceleration is defined as
  \begin{align}
    a_\text{rad}(r) & = \frac{4\pi}{c} \frac{1}{\rho(r)}  \int\limits_{0}^{\infty} \kappa_\nu H_\nu \mathrm{d}\nu ,\\
                                & = \frac{16\pi^2}{c} \frac{r^2 \varv(r)}{\dot{M}} \int\limits_{0}^{\infty} \kappa_\nu H_\nu \mathrm{d}\nu, 
  \end{align}   
        there is a leading dependence with $\dot{M}^{-1}$, therefore making $f^{-1}$ the more interesting quantity for the plots. The
        lower panel of Fig.\,\ref{fig:aradresponse} also illustrates that the particular value of $f$ changes the amplitude, but not
        the general behavior of the response, since one can scale all the results almost perfectly to the same curve $\textsc{resp}(r)$ using 
        the relation
        \begin{equation}
          \label{eq:response}
          \textsc{resp}(r) = \frac{1}{1-f} \left[ 1 - \frac{a_\text{rad}(\dot{M})}{ f \cdot a_\text{rad}(f \cdot \dot{M})} \right]\text{.}
        \end{equation}
        With the help of Eq.\,(\ref{eq:response}) it would therefore be possible to implement the detailed response of $a_\text{rad}(r)$
        to suggested changes of $\dot{M}$ in order to improve the calculation of the mass-loss rate in a hydro iteration. However, similar
        to what has been discussed for the $\alpha(r)$-approach from \citet{GH2005}, this    would double the CMF calculation time before
        each update of the velocity field. Furthermore, the relation (\ref{eq:response}) cannot account for the typically more complex 
        changes of $\varv(r)$ and thus even a mass-loss rate obtained in this detailed way does not lead to a better model convergence. 
        In fact, such methods have been tested and have turned out not to be better than the more approximate way described below.
        
\begin{figure}[ht]
  \resizebox{\hsize}{!}{\includegraphics{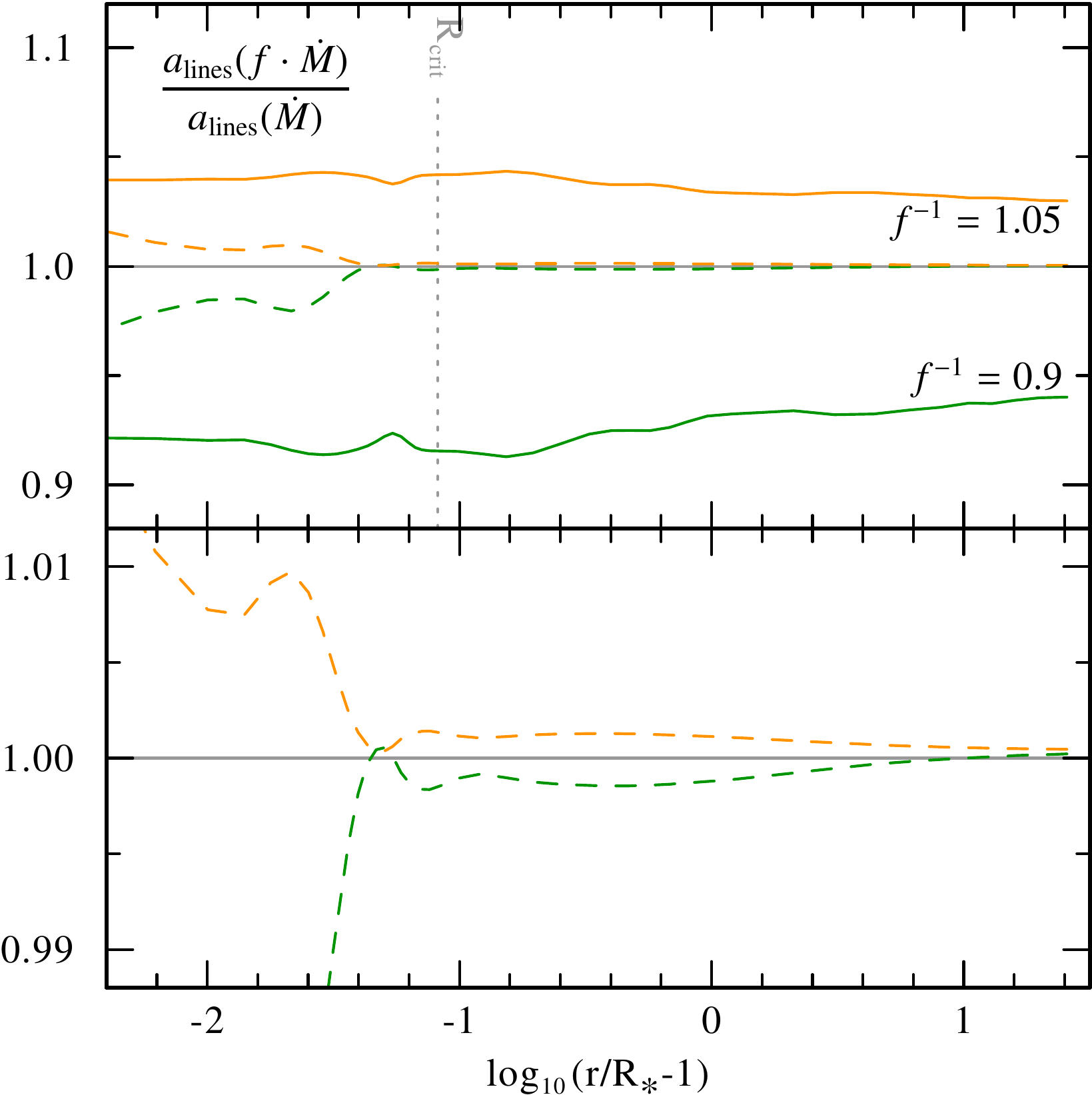}}
  \caption{Upper panel: Response of the radiative line (solid curves) and continuum (dashed curves) acceleration for two different 
                 change factors $f$ of the mass-loss rate $\dot{M}$. Similar to Fig.\,\ref{fig:aradresponse}, the ratio of the modified 
                                         to unmodified acceleration is plotted.
                                         Lower panel: Zoom-in of the curves showing the continuum response.}
  \label{fig:respapprox}
\end{figure}

  Apart from calculating the total response of the radiative acceleration in our test calculations, we also calculated the isolated
        response of the line and the total continuum term. In Fig.\,\ref{fig:respapprox} the results for two different mass-loss 
        modification factors $f$ are shown. As we can see, the continuum shows only a very small response that can be neglected, while the line 
        response is stronger and roughly of the order $f^{-1}$. In fact $f^{-1}$ is never completely reached, especially not in the outer wind,
        but since we want to use our approximation only for the calculation of the mass-loss rate, this is not a problem. By slightly
        over-predicting the effect of the change in $\dot{M}$, we avoid potential ``overshooting'' of the correction. We therefore assume for
        our calculations of the mass-loss rate update, that the
        radiative acceleration $a_\text{rad}$ changes for a mass-loss rate modified by the factor $f$ such that    
\begin{equation}
  a_\text{rad}(f \cdot \dot{M}) = \frac{1}{f} a_\text{lines}(\dot{M}) + a_\text{cont}(\dot{M}) \text{,}
\end{equation}  
  with $a_\text{rad}(\dot{M}) = a_\text{lines}(\dot{M}) + a_\text{cont}(\dot{M})$ denoting the acceleration with the original mass-loss
        rate. Using this assumption, we calculate the factor $f$ which would be necessary to obtain $\mathcal{\tilde{F}} = 0$ at the current
        $r_\mathcal{\tilde{G}}$. For a converged model, this is automatically fulfilled and we obtain $f = 1$, that is, no more change in $\dot{M}$.
        In the general case, the condition $\mathcal{\tilde{F}}(f\,\dot{M}) \stackrel{!}{=} 0$ leads to
\begin{equation}
  \label{eq:fmodmdot}
  f = \frac{\Gamma_\text{lines}(r_\mathcal{\tilde{G}})}
                 { 1 -  \Gamma_\text{cont}(r_\mathcal{\tilde{G}}) - 2 \frac{a(r_\mathcal{\tilde{G}}) \cdot r_\mathcal{\tilde{G}}}{GM} 
                                                  \left[ a(r_\mathcal{\tilde{G}}) - r_\mathcal{\tilde{G}} \left.\frac{\mathrm{d}a}{\mathrm{d}r}\right|_{r=r_\mathcal{\tilde{G}}} \right]}
,\end{equation}
  with $\Gamma_\text{lines} = a_\text{lines}/g$ and $\Gamma_\text{cont} = a_\text{cont}/g$. To avoid overly large corrections, which could
        significantly disturb the model convergence, only 50\% of the calculated change for $\dot{M}$ is usually applied in one iteration.
        
        The described update of the mass-loss rate now leads to a convergence of $r_\mathcal{\tilde{F}}$ and $r_\mathcal{\tilde{G}}$.
        However, there is so far no guarantee that the total optical depth of the converged model will be identical to that of the starting model. To
        ensure also the latter, the whole atmosphere stratification is radially adjusted before integrating the hydrodynamic equation. This adjustment
        is performed already with the new mass-loss rate and ensures the conservation of the total Rosseland continuum optical depth.

 \subsection{Scheme of the hydrodynamic stratification update}  
    \label{sec:hdscheme}
  
        The implementation concept of the hydrodynamical stratification update in regards to the overall model iteration is similar to the
        one described in \citet{Sander+2015} for models with a consistent quasi-hydrostatic part. In fact, the full hydrodynamic treatment
        and quasi-hydrostatic update are alternative branches for the step (2d) in the overall iteration scheme outlined in Sect.\,\ref{sec:itscheme}.
        In order to ensure that the overall model calculations are not vastly disrupted by a stratification update, such updates are 
        not performed during each iteration, but only immediately before the Eddington factors in the following radiative transfer job are recalculated
        and if the overall corrections to the population numbers are below a certain prespecified level (in addition, it is also possible to
        require a certain level of flux consistency for a stratification update).
        
        The hydrodynamic stratification update itself can be described by the following scheme:
        
        \begin{enumerate}
          \item Check whether the hydrodynamic equation is fulfilled at all depth points: If so, no update needs to be performed.
                \item The quantities $\Gamma_\text{rad}(r)$ and $a(r)$ are calculated based on the current model stratification. 
                \item If $r_\mathcal{\tilde{F}} \neq r_\mathcal{\tilde{G}}$, the mass loss rate $\dot{M}$ is updated by a factor $f$ as described in Eq.\,(\ref{eq:fmodmdot})
                \item Iteration:
                  \begin{enumerate}
                          \item Starting from $\varv(r_\mathcal{\tilde{F}}) = a(r_\mathcal{\tilde{F}})$, the new velocity field is obtained via integrating Eq.\,(\ref{eq:hdfg})
                                with a fourth-order Kunge-Kutta method using adaptive step sizes. The quantities $\mathcal{F}$ and $\mathcal{G}$ are calculated on the fly.  
                                To avoid numerical issues near the critical point, we make use of l'H\^{o}pital's rule. As the hydrodynamic integration requires a resolution below
                                the regular depth grid, we use spline interpolation to obtain the interstice values for $\Gamma_\text{rad}$ and $a$. 
                                \item With the new $\varv(r)$ now given, we calculate the resulting new density stratification and total Rosseland continuum optical depth $\tau_\text{Ross}(R_\ast)$
                                \item If the $\tau_\text{Ross}(R_\ast)$ is conserved, the iteration ends. Otherwise $\Gamma_\text{rad}(r)$ and $a(r)$ are shifted radially and 
                                      the next iteration cycle is started.
                  \end{enumerate}
          \item If necessary, the depth grid spacing is updated to better reflect the new stratification. All necessary quantities are
                      interpolated from the old to the new grid.
        \end{enumerate}
        
        After the hydrodynamic stratification update is complete, the overall iteration cycle continues with the next radiative transfer calculation (we refer also to Sect.\,\ref{sec:itscheme} for
        details of the overall iteration). Due to the inclusion in the overall iteration cycle, the values of $\Gamma_\text{rad}(r)$ and $a(r)$ used in the next hydrodynamic stratification
        update implicitly include all effects of the former stratification and mass-loss rate update. While this procedure can significantly increase the total number of overall iterations
        compared to non-HD models, this iterative approach allows us to refrain from any (semi-)analytical assumptions for the radiative acceleration, making this method essentially
        applicable for the whole range of stars that can be described by PoWR atmosphere models.
        The atmosphere model is eventually considered to be converged if all of the following requirements are met:
        \begin{itemize}
          \item The relative corrections to the population numbers are below a certain level (typically $10^{-3}$).
                \item Flux consistency is achieved within a certain accuracy (typically setting: relative departures may not be larger than $10^{-2}$).
                \item The hydrodynamical equation is fulfilled throughout the whole atmosphere 
                      within a specified accuracy (typically $5\%$, but we allow larger deviations 
                                        at the inner and outer boundary).
        \end{itemize}
        With the exception of the last point of course, these criteria are the same as for non-HD models, which are used for reproducing observed 
        spectra and empirically obtain stellar and wind parameters. Based on the converged atmosphere model, the emergent spectrum is subsequently 
        calculated in the observer's frame, allowing us to cross-check the results with observed spectra.

\section{Results}
  \label{sec:results}
  
  In order to test whether or not our new method is applicable to OB stars, where the approach from \citet{GH2008} failed, we
        calculated a hydrodynamically consistent model for the well-studied O supergiant $\zeta$\,Pup/\object{HD\,66811}. Starting from the parameters given in \citet{Bouret+2012},
        we first calculated a standard PoWR model using a prescribed $\beta$-law connected to a consistent quasi-hydrostatic part as described in
        \citet{Sander+2015}. While a model reproducing most spectral features already requires Ne, Mg, Si, P, and S to be considered, the work ratio
        of such a model is only $Q = 0.74$. By adding further elements, most notably Ar, the work ratio was close to unity and the model could
        be used as a starting approach for the hydrodynamic calculations.

\begin{table}
  \caption{Input parameters for the $\zeta$\,Pup model}
  \label{tab:modelinput}
  \centering
  \begin{tabular}{l c c c}
  \hline\hline
           Parameter                                            &  \multicolumn{3}{c}{Value}     \\
        \hline
    $T_*$\,[kK]  \rule[0mm]{0mm}{3mm}                     & \multicolumn{3}{c}{  $42.0$ }  \\
    $R_*$\,[$R_\odot$]                                    & \multicolumn{3}{c}{  $15.9$ }  \\
    $\log L$\,[$L_\odot$]                                 & \multicolumn{3}{c}{  $5.85$ }  \\
    $M_\ast$\,[$M_\odot$]                                 & \multicolumn{3}{c}{    $45$ }  \\
    $\log g$\,[cm\,s$^{-2}$]                              & \multicolumn{3}{c}{   $3.7$ }  \\
    $D_\infty$                                            & \multicolumn{3}{c}{    $10$ }  \\
    \medskip
    $\varv_\text{mic}$\,[km/s]                            & \multicolumn{3}{c}{    $15$ }  \\                 
    \textit{abundances}                                   & \textit{mass fractions}        \\
    $X_\text{H}$\tablefootmark{a} \rule[0mm]{0mm}{3mm}    & \multicolumn{3}{c}{$0.6$}      \\
    $X_\text{He}$\tablefootmark{a}                        & \multicolumn{3}{c}{$0.383$}    \\
    $X_\text{C}$\tablefootmark{a}                         & \multicolumn{3}{c}{$2.86 \times 10^{-4}$} \\
    $X_\text{N}$\tablefootmark{a}                         & \multicolumn{3}{c}{$1.05 \times 10^{-2}$} \\
    $X_\text{O}$\tablefootmark{a}                         & \multicolumn{3}{c}{$1.30 \times 10^{-3}$} \\
    $X_\text{Ne}$\tablefootmark{b}                        & \multicolumn{3}{c}{$1.26 \times 10^{-3}$} \\
    $X_\text{Mg}$\tablefootmark{b}                        & \multicolumn{3}{c}{$6.92 \times 10^{-4}$} \\
    $X_\text{Si}$\tablefootmark{b}                        & \multicolumn{3}{c}{$0.70 \times 10^{-3}$} \\
    $X_\text{P}$\tablefootmark{b}                         & \multicolumn{3}{c}{$6.15 \times 10^{-6}$} \\
    $X_\text{S}$\tablefootmark{b}                         & \multicolumn{3}{c}{$3.09 \times 10^{-4}$} \\
    $X_\text{Cl}$\tablefootmark{b}                        & \multicolumn{3}{c}{$8.20 \times 10^{-6}$} \\
    $X_\text{Ar}$\tablefootmark{b}                        & \multicolumn{3}{c}{$7.34 \times 10^{-5}$} \\
    $X_\text{K}$\tablefootmark{b}                         & \multicolumn{3}{c}{$3.14 \times 10^{-6}$} \\
    $X_\text{Ca}$\tablefootmark{b}                        & \multicolumn{3}{c}{$6.13 \times 10^{-5}$} \\
    $X_\text{Fe}$\tablefootmark{b,c}                      & \multicolumn{3}{c}{$1.40 \times 10^{-3}$} \\
            
  \hline
  \end{tabular}
  \tablefoot{
        \tablefoottext{a}{Abundance taken from \citet{Bouret+2012}}
        \tablefoottext{b}{Solar abundances, taken from \citet{Asplund+2009}}
        \tablefoottext{c}{Fe include also the further iron group elements Sc, Ti, V, Cr, Mn, Co, and Ni.
                          See \citet{GKH2002} for relative abundances.}
  }  
\end{table}             
                
        In our first approach, we applied the same clumping stratification as \citet{Bouret+2012}, that is, depth-dependent \mbox{(micro-)}clumping 
        with a maximum value of $D_\infty = 20$ or $f_{\text{V},\infty} = D^{-1}_\infty = 0.05$ and no interclump medium. This is 
        a standard approach in state-of-the-art atmosphere models for hot and massive stars \citep[e.g.,][]{HK1998,HM1999} and allows one to calculate the population numbers for the 
        clumped wind, which has a density increased by a factor $D(r)$ compared to a smooth wind. For the radiative transfer, on the other hand, one can average
  between the clump and interclump medium as clumps are assumed to have a small size in comparison to the mean-free path of the photons.
        Furthermore, instead of the 
        standard description of depth-dependent clumping  in PoWR \citep{GH2005}, we employed the same parametrization as in the CMFGEN model 
        from \citet{Bouret+2012},
        namely
        \begin{equation}
          f_\text{V}(r) = f_{\text{V},\infty} + (1 -  f_{\text{V},\infty}) \cdot \exp\left(-\frac{\varv(r)}{\varv_\text{cl}}\right)
        ,\end{equation}
        introduced in \citet{Martins+2009}, where the clumping ``onset'' is described by a velocity $\varv_\text{cl}$. In their analysis for $\zeta$\,Pup,
        \citet{Bouret+2012} use $\varv_\text{cl} = 100\,$km/s. We started with a similar stratification, but quickly realized
        that the value for $\varv_\text{cl}$ is not sufficient and leads to solutions with an overly high terminal velocity together with an overly low
        mass-loss rate. Stratifications where we set $\varv_\text{cl} = 0.5\,\varv_\text{sonic}$ lead to better results, but since the sonic
        point can change during the iterations, we decided to implement another clumping stratification with
        \begin{equation}
          f_\text{V}(r) = f_{\text{V},\infty} + (1 -  f_{\text{V},\infty}) \cdot \exp\left(-\frac{\tau_\text{cl}}{\tau_\text{Ross}(r)}\right)\text{,}
        \end{equation}
        that is, where we specify the clumping onset via an optical depth $\tau_\text{cl}$ instead of a particular velocity. This approach was eventually applied in the
        final model presented in this work. As we discuss later on, it was furthermore necessary to reduce the maximum clumping value to $D_\infty = 10$
        in our model. The complete set of input parameters for the final hydrodynamical model is compiled in Table\,\ref{tab:modelinput}.

\begin{table}
  \caption{Atomic data used in the hydrodynamic model}
  \label{tab:datom}
  \centering
  \begin{tabular}{l r r p{3mm} l r r}
  \hline\hline
    Ion  \rule[0mm]{0mm}{3mm}                &   Levels &   Lines\tablefootmark{a}   &    &
                Ion  \rule[0mm]{0mm}{3mm}                &   Levels &   Lines\tablefootmark{a}  \\
  \hline

    \ion{H}{i}  \rule[0mm]{0mm}{3mm} &    10  &    45     &  &     \ion{P}{iv}    &    12     &      16     \\         
    \ion{H}{ii}                      &     1  &     0     &  &     \ion{P}{v}     &    11     &      22     \\                
    \ion{He}{i}                      &    17  &    55     &  &     \ion{P}{vi}    &     1     &       0     \\         
    \ion{He}{ii}                     &    16  &   120     &  &     \ion{S}{iv}    &    11     &      14     \\         
    \ion{He}{iii}                    &     1  &     0     &  &     \ion{S}{v}     &    10     &      13     \\                
    \ion{C}{ii}                      &    32  &   148     &  &     \ion{S}{vi}    &    22     &      75     \\         
    \ion{C}{iii}                     &    40  &   226     &  &           \ion{Cl}{iii}  &     1     &       0     \\     
    \ion{C}{iv}                      &    25  &   230     &  &     \ion{Cl}{iv}   &    24     &      34     \\                  
    \ion{C}{v}                       &    29  &   120     &  &     \ion{Cl}{v}    &    18     &      29     \\         
    \ion{C}{vi}                      &     1  &     0     &  &     \ion{Cl}{vi}   &    23     &      46     \\          
    \ion{N}{ii}                      &    38  &   201     &  &     \ion{Ar}{ii}   &    20     &      33     \\
    \ion{N}{iii}                     &    87  &   507     &  &     \ion{Ar}{iii}  &    14     &      13     \\                           
    \ion{N}{iv}                      &    38  &   154     &  &     \ion{Ar}{iv}   &    13     &      20     \\          
    \ion{N}{v}                       &    20  &   114     &  &     \ion{Ar}{v}    &    10     &      11     \\         
    \ion{N}{vi}                      &    14  &    48     &  &     \ion{Ar}{vi}   &     9     &      11     \\  
    \ion{N}{vii}                     &     2  &     1     &  &     \ion{Ar}{vii}  &    20     &      34     \\                   
    \ion{O}{ii}                      &    37  &   150     &  &     \ion{Ar}{viii} &    11     &      24     \\    
    \ion{O}{iii}                     &    33  &   121     &  &           \ion{Ar}{ix}   &    10     &      10     \\             
    \ion{O}{iv}                      &    29  &    76     &  &     \ion{Ar}{x}    &     3     &       1     \\         
    \ion{O}{v}                       &    36  &   153     &  &     \ion{K}{iii}   &    20     &      40     \\          
    \ion{O}{vi}                      &    16  &   101     &  &     \ion{K}{iv}    &    23     &      27     \\         
    \ion{O}{vii}                     &     1  &     0     &  &     \ion{K}{v}     &    19     &      33     \\                
    \ion{Ne}{i}                      &     8  &    14     &  &     \ion{K}{vi}    &    28     &      38     \\         
    \ion{Ne}{ii}                     &    18  &    40     &  &     \ion{K}{vii}   &     1     &       0     \\          
    \ion{Ne}{iii}                    &    18  &    18     &  &     \ion{Ca}{iv}   &    24     &      43     \\          
    \ion{Ne}{iv}                     &    35  &   159     &  &     \ion{Ca}{v}    &    15     &      12     \\                 
    \ion{Ne}{v}                      &    25  &    37     &  &     \ion{Ca}{vi}   &    15     &      17     \\
    \ion{Ne}{vi}                     &    25  &    53     &  &     \ion{Ca}{vii}  &    20     &      28     \\
    \ion{Ne}{vii}                    &    25  &    61     &  &     \ion{Fe\tablefootmark{b}}{ii}    &     1     &       0     \\  
    \ion{Ne}{viii}                   &    25  &   101     &  &     \ion{Fe\tablefootmark{b}}{iii}   &    13     &      40     \\   
    \ion{Ne}{ix}                     &    23  &    51     &  &     \ion{Fe\tablefootmark{b}}{iv}    &    18     &      77     \\  
    \ion{Mg}{ii}                     &     1  &     0     &  &     \ion{Fe\tablefootmark{b}}{v}     &    22     &     107     \\        
    \ion{Mg}{iii}                    &    11  &    16     &  &     \ion{Fe\tablefootmark{b}}{vi}    &    29     &     194     \\  
    \ion{Mg}{iv}                     &    10  &     9     &  &     \ion{Fe\tablefootmark{b}}{vii}   &    19     &      87     \\   
    \ion{Mg}{v}                      &    10  &     9     &  &     \ion{Fe\tablefootmark{b}}{viii}  &    14     &      49     \\   
    \ion{Mg}{vi}                     &    10  &    19     &  &     \ion{Fe\tablefootmark{b}}{ix}    &    15     &      56     \\ 
    \ion{Mg}{vii}                    &    10  &    11     &  &     \ion{Fe\tablefootmark{b}}{x}     &     1     &       0     \\         
                \ion{Mg}{viii}                   &     1  &     0     &  &                                &           &             \\    
    \ion{Si}{iii}                    &    24  &    69     &  &                         &           &             \\
    \ion{Si}{iv}                     &    55  &   465     &  &                     &           &             \\
    \ion{Si}{v}                      &    52  &   265     &  &                   Total  &    1450   &    5221     \\      
  \hline
  \end{tabular}
  \tablefoot{
        \tablefoottext{a}{The number of lines refers to those transitions that have non-negligible oscillator strengths and 
                                                  are therefore considered in the radiative transfer calculation.}
        \tablefoottext{b}{For Fe, the number of levels and lines refer to superlevels and superlines.
                                                  Fe includes also the further iron group elements Sc, Ti, V, Cr, Mn, Co, and Ni.
                          See \citet{GKH2002} for concept details and relative abundances.}
  }  
\end{table}

In models with a predefined velocity law in the wind part, it is usually sufficient to include
only those elements that can either be seen in the spectrum or contribute significantly to the blanketing. However,
for the hydrodynamic models, it is essential to include all ions that have a significant contribution to the radiative 
force, even if they neither leave a noticable imprint in the spectrum nor significantly affect the blanketing. Yet,
accounting for all elements and their ions from hydrogen up to the iron group would be numerically extremely 
expensive and thus practically impossible. Fortunately, various elements and ions are only important in a certain parameter regime 
and thus can be neglected outside of these. A list of the ions considered in the model presented in this work is given in Table\,\ref{tab:datom}.

\begin{figure}[ht]
  \resizebox{\hsize}{!}{\includegraphics{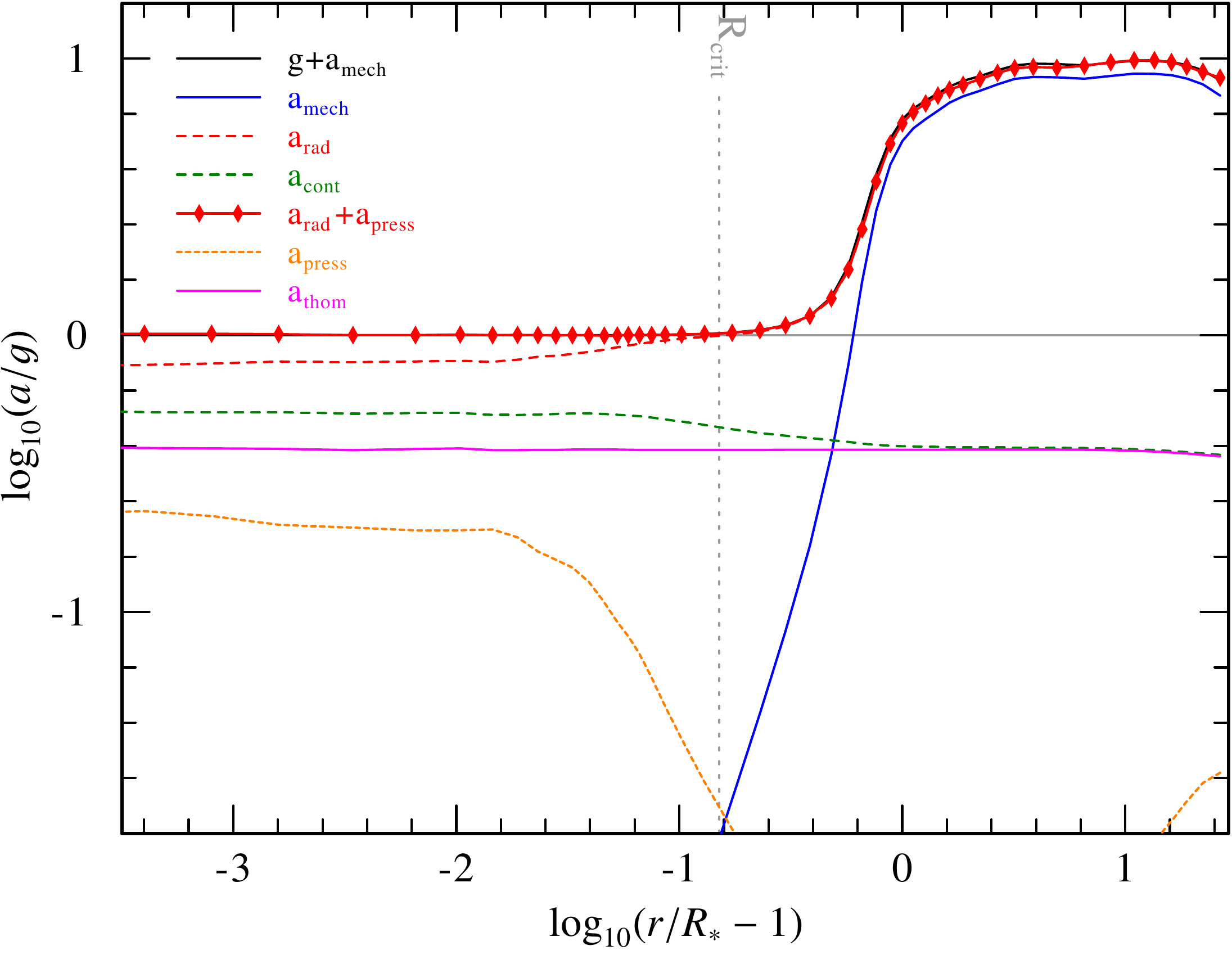}}
  \caption{Acceleration stratification for a hydrodynamically consistent 
           model. The wind acceleration (thick red diamond line) is compared to
           the repulsive sum of inertia and gravitational acceleration $g(r)$ (black
           line). The fact that these two curves are (almost) identical 
                                         illustrates that the hydrodynamic equation is fulfilled throughout the stellar
                                         atmosphere. For a more convenient illustration, 
                                         all terms are normalized to $g(r)$. The input parameters of
                                         the model are compiled in Table\,\ref{tab:modelinput} while the resulting
                                         quantities can be found in Table\,\ref{tab:modelparam}.}
  \label{fig:zpupacc}
\end{figure}

The acceleration balance of the hydrodynamically consistent model is shown in Fig.\,\ref{fig:zpupacc}.
Throughout the atmosphere, an excellent agreement between the outward and inward forces is obtained.
Up until the critical point, not only the line acceleration and the Thomson term are important,
but there are also significant contributions from the gas pressure and the true continuum, that is,
the continuum not produced by Thomson scattering, to the driving.
In the wind part, both of the latter terms become negligible. However, it has to be noted that this
is not necessarily the case for all kinds of hot stars. In the more dense Wolf-Rayet winds, situations
can occur where the true continuum is not negligible in the wind \citep[see, e.g., the consistent model 
for WR\,111 in][]{GH2005}. The importance of the pressure 
term strongly depends on the assumptions for microturbulence. In this model, a constant value of 
$\varv_\text{mic} = 15\,$km/s was used in the hydrodynamic calculations. When using larger
values or depth-dependent descriptions with $\varv_\text{mic}$ increasing outwards, the 
$a_\text{press}$-term can become significant again in the outer wind.

\begin{figure}[ht]
  \resizebox{\hsize}{!}{\includegraphics{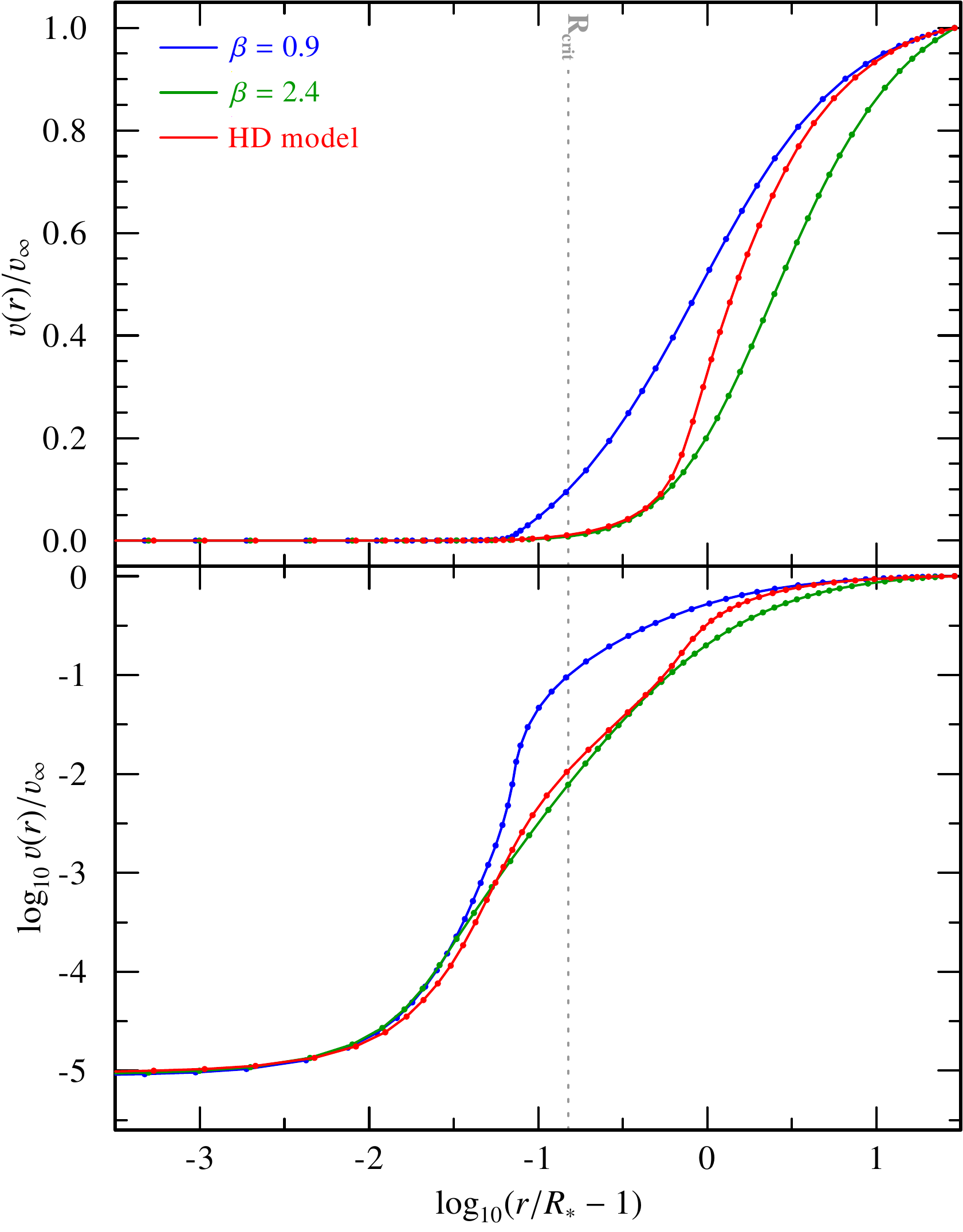}}
  \caption{Normalized velocity field for the hydrodynamically-consistent model (red) versus
                 a model using a $\beta$-law connected to a consistent quasi-hydrostatic
                                         part. The upper panel shows the velocity in non-logarithmic units, thereby highlighting 
                                         the wind part, while the lower panel displays the normalized velocity in logarithmic
                                         units, thus focusing on the inner layers.}
  \label{fig:zpupvelo}
\end{figure}

\begin{figure*}[p!]
  \includegraphics[angle=0,width=0.95\textwidth]{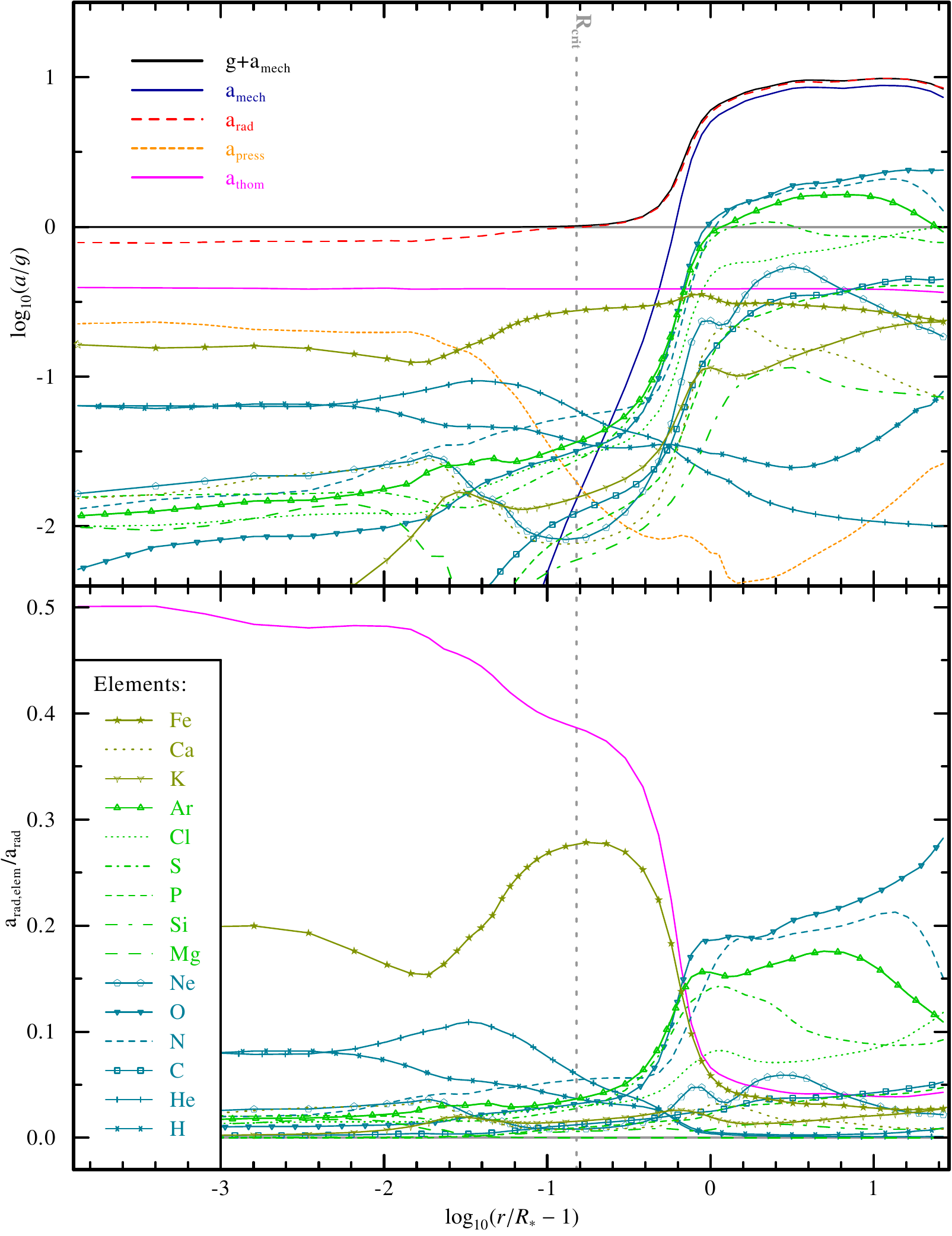}
  \caption{Absolute (upper panel) and relative (lower panel) contributions to the radiative
                 acceleration from the different elements considered in the hydrodynamically consistent
                                         atmosphere model. The total radiative acceleration and the acceleration 
                                         due to gas pressure are also shown in the upper panel for comparison. The lower panel shows the extent to which electron scattering (pink solid curve) and the various elements contribute to the 
                                         total radiative acceleration.}
  \label{fig:zpupelem}
\end{figure*}

The resulting velocity field for the converged hydrodynamic models is shown in Fig.\,\ref{fig:zpupvelo},
where it is compared to the stratification of two standard models using a prescribed velocity field 
in the form of so-called $\beta$-laws, that is, 
\begin{equation}
  \label{eq:betasimple}
        \varv(r) = \varv_\infty \left( 1 - \frac{R_\ast}{r} \right)^{\beta}\text{.}
\end{equation}
When implementing the $\beta$-law into a stellar atmosphere model, Eq.\,(\ref{eq:betasimple}) is usually
slightly modified due to several reasons, most prominently the necessity to connect the wind domain with
a proper quasi-hydrostatic domain \citep[see][for more details]{Sander+2015} and the numerical issues that
would occur for $\varv(R_\ast) = 0$. This modification can be done in more than one way and thus
also differ between different stellar atmosphere codes. In PoWR, two ways of ``fine parametrization'' for
the $\beta$-law are available, namely
\begin{equation}
  \label{eq:betafineshift}
        \varv(r) = p \left( 1 - \frac{R_\ast}{r + R_\text{s}} \right)^{\beta}
\end{equation}
and
\begin{equation}
  \label{eq:betafinescale}
        \varv(r) = p \left( 1 - f_\text{s} \frac{R_\ast}{r} \right)^{\beta}
,\end{equation}
with their fine parameters $p$ and $R_\text{s}$ or $f_\text{s}$, respectively. While $R_\text{s}$ or $f_\text{s}$
are responsible for a proper connection of the wind and the quasi-hydrostatic domain, $p \approx \varv_\infty$ ensures that the
specified terminal velocity is reached at the outer boundary. All fine parameters are automatically calculated
depending on the choices of the velocity field, the connection criterion and whether the parametrization from
Eq.\,(\ref{eq:betafineshift}) or from Eq.\,(\ref{eq:betafinescale}) should be used. 

For the hydrodynamically consistent model presented in this work, there is of course no prescribed wind velocity
field, but for the comparison calculations we had to make a choice and used Eq.\,(\ref{eq:betafineshift}) since
it is more widely used in modern PoWR models. 
Interestingly the velocity field in the lower part of the wind, just above
the sonic point, can be approximated with a beta law using $\beta = 2.4$, but the outer part of the wind 
is best matched with $\beta = 0.9$. Due to the fact that the fine parametrization is not unique as described above, these
deduced $\beta$-law approximations can vary slightly ($10$ to $20$\%) when using different 
fine parametrizations.
In-between the two parts there is a steep increase of the velocity, steeper than could be modeled
by a $\beta$-law connected to the quasi-hydrostatic part.
The reasons for this kind of velocity field are revealed when looking at the particular contributions to the radiative acceleration
plotted in Fig.\,\ref{fig:zpupelem}. Around and shortly above the critical point, only the iron group elements 
and the electron scattering contribute significantly to the radiative acceleration. In contrast, further outwards, 
many more elements contribute, and for $r > 1.6\,R_\ast$ N, O, and Ar start to exceed not only $\Gamma_\text{e}$,
but also the contribution of the iron group elements. S, Cl, and Ne follow further out and for 
$r \gtrsim 2\,R_\ast$ also the contributions of C and P are comparable to $\Gamma_\text{e}$, having
even a bit more impact than the iron group at this distance.

The complex contribution to the radiative force from the various elements is directly imprinted in the
resulting velocity field and thus ``naturally'' explains the deviations from a standard $\beta$-law which has
been derived from a taylored fit of the UV resonance line profiles already by \citet{Hamann1980}. While
we do not see a noticeable velocity plateau in the final model, such a plateau occurred in several of the test models
derived during the preparation of this work. The plateau becomes visible if the iron group contribution
already exceeds $\Gamma_\text{e}$ around or even below the sonic point, which can already happen for slightly
higher mass-loss rates than derived for $\zeta$\,Pup here. 

\begin{figure*}[p!]
  \centering
    \includegraphics[angle=0,width=0.95\textwidth]{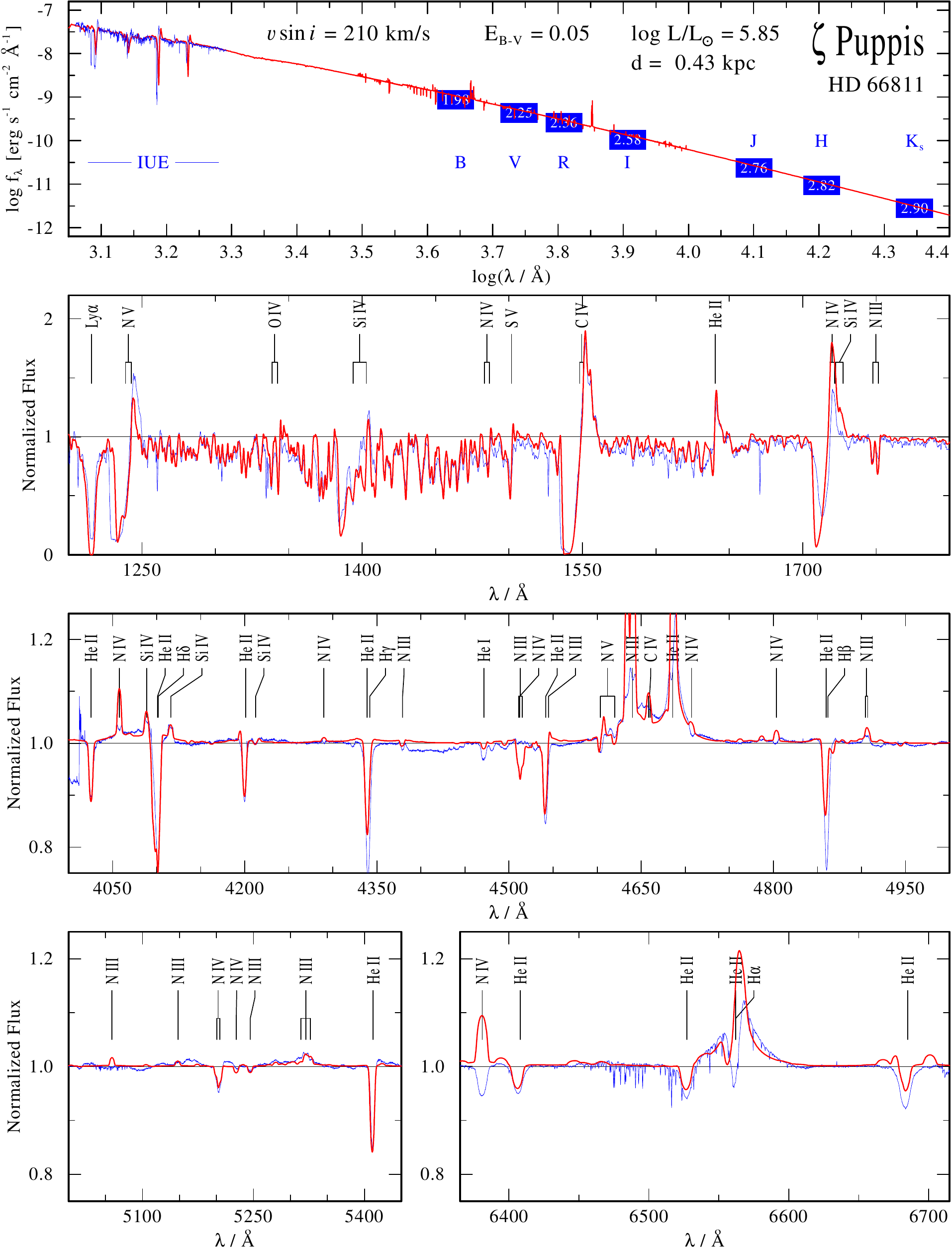}
  \caption{Comparison of the hydrodynamically consistent O-star model with observed spectra of
                 the O4 supergiant $\zeta$\,Pup. The uppermost panel shows the spectral energy distribution
                                         while the other panels compare the normalized line spectra between the model (red) and
                                         the observations (blue) in various wavelength ranges.}
  \label{fig:zpupmaster}
\end{figure*}

\begin{table}
  \caption{Results from the hydrodynamically consistent $\zeta$\,Pup model with input parameters as in Table\,\ref{tab:modelinput}}
  \label{tab:modelparam}
  \centering
  \begin{tabular}{l c c c}
  \hline\hline
           Quantity                                             &  \multicolumn{3}{c}{Value}     \\
        \hline
    $\log R_\text{t}$\,[$R_\odot$]  \rule[0mm]{0mm}{3mm}  & \multicolumn{3}{c}{  $2.00$ }  \\
    $\log Q_\text{ws}$\,[cgs]                             & \multicolumn{3}{c}{ $-12.1$ }  \\
    $T_{2/3}$\,[kK]                                       & \multicolumn{3}{c}{  $40.7$ }  \\
    $q_\text{ion}$                                            & \multicolumn{3}{c}{  $0.79$ }  \\
    $\Gamma_\text{e}$                                     & \multicolumn{3}{c}{  $0.38$ }  \\
    $\overline{\Gamma}_\text{rad}$\tablefootmark{a}       & \multicolumn{3}{c}{  $0.77$ }  \\
    \medskip
    $\log g_\text{eff}$\,[cm\,s$^{-2}$]\tablefootmark{a}  & \multicolumn{3}{c}{  $3.63$ }  \\

    $r_\text{c}$\,[$R_\ast$]                              & \multicolumn{3}{c}{  $1.16$ }  \\
    \textbf{$\mathbf{\boldsymbol{\log}\,\dot{M}}$\,[$\boldsymbol{M_\odot}\,\text{yr}^{-1}$]}  & \multicolumn{3}{c}{ $\mathbf{-5.80}$ }  \\
    \medskip
    \textbf{$\boldsymbol{\varv_\infty}$\,[km/s]}                       & \multicolumn{3}{c}{ $\mathbf{2046}$  }  \\                 
    $\eta = \dot{M} \varv_\infty c / L$                   & \multicolumn{3}{c}{  $0.23$ }  \\                 
    $\log D_\text{mom}$\tablefootmark{b} [g\,cm\,s$^{-2}$\,R$_\odot^{-1/2}$] 
                                                                                                                      & \multicolumn{3}{c}{  $29.4$ }  \\                 
  \hline
  \end{tabular}
  \tablefoot{
        \tablefoottext{a}{Effective gravity calculated via $\overline{\Gamma}_\text{rad}$ as described in \citet{Sander+2015}}
                                \tablefoottext{b}{Modified wind momentum, defined as $D_\text{mom} = \dot{M} \varv_\infty \sqrt{R_\ast/R_\odot}$ \citep[see, e.g.,][]{KP2000}}
  }  
\end{table} 

The derived parameters of our hydrodynamically consistent model are compiled in Table\,\ref{tab:modelparam},
while the spectral energy distribution and important parts of the normalized spectrum are compared to
observations in Fig.\,\ref{fig:zpupmaster}. The UV observation was obtained with the IUE satellite (SWP15296) 
while the optical spectrum stems from an earlier observation within our group (Hamann, priv. comm.). The
photometric data used in the SED plot have been taken from \cite{Ducati2002}.
While the displayed model is rather to demonstrate the new technique described in the previous
section and therefore has not been fine-tuned to precisely reproduce the spectral features 
of $\zeta$\,Pup, one can still compare the main parameters to spectral analyses. While the starting
parameters were motivated by the non-hydrodynamical model results from \citet{Bouret+2012}, their
high clumping factor of $20$ would lead to an underprediction of the H$\alpha$ electron scattering wings
and thus was reduced to the more typical value of $10$. The mass-loss rate of $\log \dot{M} = -5.8$
is slightly lower than in \citet{Bouret+2012}, but higher than the value obtained by \citet{Pauldrach+2012}. 

The emergent spectrum of our hydrodynamical model shows all the typical features of an early Of-type star
with a fast wind and strong emission in both \ion{N}{iii} $\lambda 4634$-$40$-$42$ and \ion{He}{ii} $\lambda 4686$ \citep[see, e.g.,][for a classification scheme]{Sota+2011}. 
When comparing the detailed spectral appearance to the observation of $\zeta$\,Pup in 
Fig.\,\ref{fig:zpupmaster}, one can conclude that the UV spectrum, including the iron forest, is well reproduced
apart from the precise shape of the nitrogen profiles, which can be affected by various parameters including
so-called ``superionization'' due to X-rays, which were not included in our model. The optical spectrum
reveals that the mass-loss rate might be slightly too high to reproduce this observation, since emission in H$\alpha$ and the other prominent emission lines appears slightly
too strong. Also H$\beta$ and H$\gamma$ seem to
be filled up by wind emission. The overall appearance, however, as well as the spectral energy distribution, are
nicely reproduced, illustrating that we have a realistic stellar atmosphere model that would require only
minor parameter adjustments to allow for a more detailed discussion. Unfortunately even these minor changes can
lead to a significant amount of calculation effort when constructing hydrodynamically consistent models, which
is why we refrain from further efforts in this introductory paper.

\section{Conclusions}
  \label{sec:conclusions}
        
In this work we constructed the first hydrodynamically consistent PoWR model for 
an O supergiant. A new method for the consistent solution of the hydrodynamic
equation, together with the solution of the statistical equations, the temperature stratification, and
the radiative transfer has been developed and successfully applied. This new technique enables us
to construct a new generation of PoWR models where the velocity field and the mass-loss rate are 
calculated consistently. 
To obtain the velocity field, the hydrodynamic equation is integrated inwards and outwards from the
critical point. Since we provide the radiative acceleration calculated in the comoving frame as
a function of radius, the critical point in our hydrodynamic equation is identical to the sonic point. 
The uniqueness of the critical point also provides the necessary condition to obtain the mass-loss rate.

As we calculate the velocity field from the hydrodynamic equation, it is mandatory
to include all ions that significantly contribute to the radiative acceleration somewhere in the 
atmosphere. In the case of our demonstration model, especially the inclusion
of Ar was crucial as it provides a major contribution to the driving in the outer wind, comparable to N and
O, although it does not leave detectable features in the spectral ranges typically observed.

In the region around the critical point, the most important line driving contribution stems from the
iron group elements. Although these elements turn out to be important
contributors throughout the whole atmosphere for our demonstration model, there are several other elements
exceeding their input in the outer part, namely N, O, S, Ar, Cl, C, P and partly Ne. Further follow-up
calculations for a wider parameter range will be necessary to shed light on details; namely,
which ions are responsible, and how this picture will change when transitioning to different
mass-loss or temperature regimes.

The obtained velocity field cannot be approximated by a $\beta$-law. 
The resulting mass-loss rate of our hydrodynamic model is in the range of what has been determined by
empirical analyses for the O4 supergiant $\zeta$\,Pup, and the resulting spectrum resembles the observed 
line spectrum and the spectral energy distribution (cf.~Fig.\,\ref{fig:zpupmaster}). Our calculations confirm that the  
value of the mass-loss rate crucially depends on the location of the critical point, which in turn reacts 
to several factors, such as the assumed microturbulence, the onset of clumping, and the Fe abundance.  
Follow-up research will therefore be required in order to study the precise influence of these and other
parameters.

\begin{acknowledgements}
  We would like to thank the anonymous referee for the fruitful suggestions that helped to improve this paper.
  We would also like to acknowledge helpful discussions with D. John Hillier.
  The first author of this work (A.S.) is supported by
  the Deutsche Forschungsgemeinschaft (DFG) under grant HA 1455/26. T.S. is grateful for financial support from the
        Leibniz Graduate School for Quantitative Spectroscopy in Astrophysics, a joint project of the Leibniz 
        Institute for Astrophysics Potsdam (AIP) and the Institute of Physics and Astronomy of the University of Potsdam.
        This research made use of the SIMBAD and VizieR databases, operated at CDS, Strasbourg, France.
\end{acknowledgements}


\bibliographystyle{aa} 
\bibliography{hydrozetapup}
   
\end{document}